\documentclass[aps,prl,reprint,superscriptaddress,amsmath,amssymb]{revtex4-2}
\usepackage{graphicx}
\usepackage{bm}
\usepackage{hyperref}
\hypersetup{colorlinks=true,
    citecolor=blue,
    linkcolor=black,
    urlcolor=blue}
\usepackage{microtype}
\usepackage{booktabs}
\usepackage{float}
\usepackage{etoolbox}
\usepackage{placeins}
\usepackage{xcolor}

\graphicspath{{figures/}{./}}
\newcommand{\safefig}[2]{%
  \IfFileExists{figures/#2.pdf}{\includegraphics[width=#1]{figures/#2.pdf}}{%
   \IfFileExists{#2.pdf}{\includegraphics[width=#1]{#2.pdf}}{%
   \fbox{\parbox[c][1.5in][c]{0.8\columnwidth}{\centering Missing image: #2.pdf}}}}}

\hypersetup{hypertexnames=false}

\newcommand{\REV}[1]{#1}
\newcommand{\REVcolor}{}

\begin{document}

\title{Connate Topological Superconductivity and Vortex-Position-Controlled Majorana Modes in the Intergrowth Superconductor Th$_2$Mo$_2$Ir$_2$Si$_4$C}
\author{Jiayang Li}
\affiliation{Anhui Provincial Key Laboratory of Low-Energy Quantum Materials and Devices, High Magnetic Field Laboratory, HFIPS, Chinese Academy of Sciences, Hefei, Anhui 230031, China}
\affiliation{Science Island Branch of Graduate School, University of Science and Technology of China, Hefei, Anhui 230026, China}
\author{Qing Wang}
\affiliation{Anhui Provincial Key Laboratory of Low-Energy Quantum Materials and Devices, High Magnetic Field Laboratory, HFIPS, Chinese Academy of Sciences, Hefei, Anhui 230031, China}
\author{Xun-Jiang Luo}
\email{xjluo@hmfl.ac.cn}
\affiliation{Anhui Provincial Key Laboratory of Low-Energy Quantum Materials and Devices, High Magnetic Field Laboratory, HFIPS, Chinese Academy of Sciences, Hefei, Anhui 230031, China}
\author{Mingliang Tian}
\affiliation{Anhui Provincial Key Laboratory of Low-Energy Quantum Materials and Devices, High Magnetic Field Laboratory, HFIPS, Chinese Academy of Sciences, Hefei, Anhui 230031, China}
\affiliation{Collaborative Innovation Center of Advanced Microstructures, Nanjing University, Nanjing 210093, China}
\author{Ning Hao}
\email{haon@hmfl.ac.cn}
\affiliation{Anhui Provincial Key Laboratory of Low-Energy Quantum Materials and Devices, High Magnetic Field Laboratory, HFIPS, Chinese Academy of Sciences, Hefei, Anhui 230031, China}
\affiliation{Collaborative Innovation Center of Advanced Microstructures, Nanjing University, Nanjing 210093, China}

\begin{abstract}
The recently synthesized intergrowth compound Th$_2$Mo$_2$Ir$_2$Si$_4$C is an experimentally established bulk superconductor whose topological electronic structure remains unexplored. Here, we identify it as a candidate for connate topological superconductivity and reveal two microscopic routes to vortex Majorana zero modes. First-principles and tight-binding-model calculations uncover Mo--Th band inversion and helical surface states enabled by interorbital spin--orbit coupling between Mo--d$_{xz/yz}$ and Th--d$_{x^2-y^2}$ orbitals. Vortex Bogoliubov--de Gennes calculations further establish a relatively robust surface-state-associated Kitaev phase and a weakly gapped, Th-orbital-derived Kitaev phase. For the latter, shifting the vortex core closes and reopens the vortex-line gap and reverses its one-dimensional topological invariant, consistent with an effective Kitaev-chain mass inversion. Finite-length-vortex calculations reveal distinct spatial and spin-resolved signatures of their Majorana end modes. The orbital Kitaev phase can also exist without helical surface states in a modified model. Although the small orbital-derived gap limits quantitative predictions of experimental observability in this compound, the uncovered mechanism suggests a route to Majorana vortices in multiorbital superconductors beyond the conventional surface-state scenario.
\end{abstract}
\maketitle

\textit{Introduction.---}
Majorana zero modes (MZMs) bound to superconducting vortices provide a promising platform for anyons and fault-tolerant quantum computation ~\cite{ReadGreen2000,Kitaev2001,Ivanov2001,Nayak2008,Alicea2012,Beenakker2013,Aguado2017,SatoAndo2017,ChiuTeoSchnyderRyu2016}. In the Fu--Kane proposal, superconducting pairing is induced in the helical surface states of a topological insulator through proximity effect~\cite{FuKane2008,PhysRevLett.114.017001}. Heterostructure realizations, however, rely on interfaces that may introduce disorder and modify the low-energy electronic structure. An alternative is connate topological superconductivity~\cite{hao2014topological, shi2017fete1, hao2019topological,YangLiHao2025}, in which helical electronic states and bulk superconductivity coexist within a single material. Fe(Te,Se) has emerged as a leading material realization of this idea\cite{hao2014topological, hao2019topological, wang2015topological, wu2016topological, wu2015cafeas, xu2016topological,ZhangWangWu2019}, with evidence for band inversion, superconducting surface states, and vortex zero-bias peaks~\cite{Zhang2018,Wang2018,Machida2019,Kong2019,Zhu2020,YinWuWang2015,LiuChenZhang2018,LiuCaoZhu2020,KongCaoZhu2021,FanYangQian2021,ChiuMachidaHuang2020}. Extending this paradigm to a stoichiometric compound with experimentally established bulk superconductivity and an additional route to vortex Majorana modes would broaden the materials landscape for
Majorana physics.

An important question beyond the Fu-Kane route is
whether a vortex-core band derived from ordinary atomic
orbitals can itself realize an effective one-dimensional Kitaev chain~\cite{Lutchyn2010,Oreg2010,NadjPerge2014,RubyPientkaPeng2015,song2022phase}. Earlier studies identified nodal and
competing vortex phases in multiband systems~\cite{qin2019quasi,hu2023topological,hu2022competing,yu2026multiband,kheirkhah2021vortex,LuoPanShiWu2025,LuoTian2026,TuWanHao2024,WangRodriguezJiao2020,JiangDaiWang2019,LuoPanLiu2021,ShiWangFuZhangHao2026},
while topological vortices from trivial normal-state bands
were also demonstrated in a multicomponent Luttinger-
semimetal model ~\cite{hu2023topological}. Here we ask whether a local
orbital-derived vortex channel can support Majorana end
modes, and whether its topology can be switched by changing the
vortex position relative to the atomic lattice. Unlike a
mechanism built around a Luttinger band touching, the
proposed route is tied to an atomic-orbital-derived core
band and need not rely on a helical surface state.

Recently, Li et al. synthesized the stoichiometric intergrowth compound Th$_2$Mo$_2$Ir$_2$Si$_4$C through block-layer design~\cite{LiJACS2025}. This compound crystallizes in the centrosymmetric tetragonal space group $I4/mmm$, with alternating ThMo$_2$Si$_2$C and ThIr$_2$Si$_2$ blocks, and exhibits bulk superconductivity below $T_c\simeq3.4$ K, as established by resistivity, magnetic susceptibility, and specific-heat measurements. Here we identify this realized superconductor
as a candidate connate platform and uncover two distinct microscopic origins of Majorana vortex states: a
surface-associated channel originating from the inverted
Mo--Th electronic sector, and a Th-orbital-derived core
channel with a much smaller protecting gap. The latter undergoes a position-driven topological transition at
fixed chemical potential, described quantitatively by an
effective one-dimensional Kitaev model. This distinct
orbital-derived mechanism motivates a broader search for
Majorana modes in multiorbital superconductors, including systems without topological surface states.

\begin{figure*}[t]
\centering
\includegraphics[width=1\textwidth]{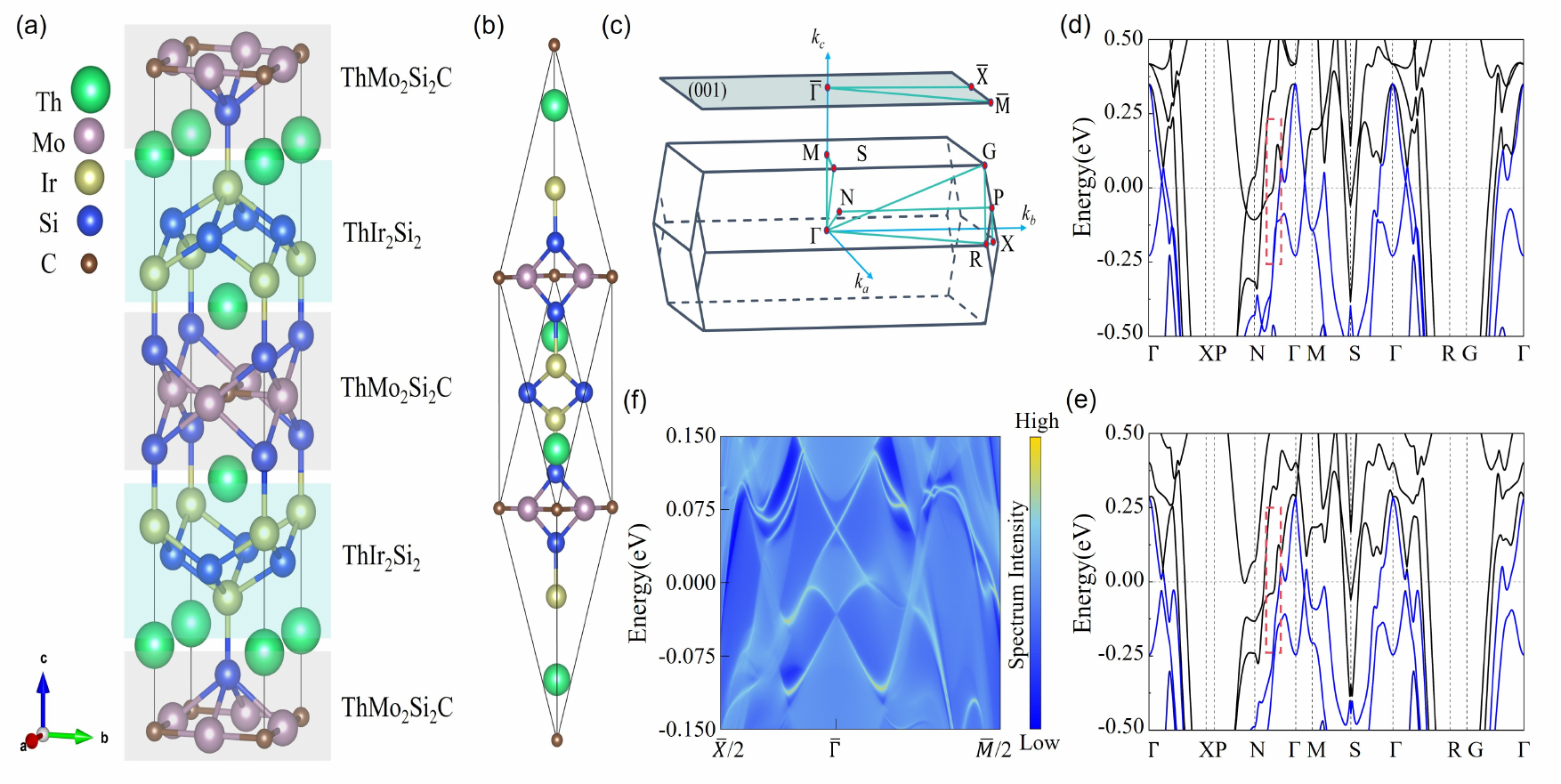}
\caption{Crystal and electronic structures of Th$_2$Mo$_2$Ir$_2$Si$_4$C. (a),(b) Intergrowth crystal structure and primitive cell of Th$_2$Mo$_2$Ir$_2$Si$_4$C. (c) Bulk and (001) surface Brillouin zones. (d),(e) Representative first-principles bulk bands without and with SOC, respectively.  (f) Calculated (001) surface spectral function exhibiting a Dirac-like surface state near $\bar\Gamma$.}
\label{fig:band}
\end{figure*}

\textit{Orbital inversion and topological surface states.---}
The intergrowth structure of Th$_2$Mo$_2$Ir$_2$Si$_4$C comprises alternating
ThMo$_2$Si$_2$C and ThIr$_2$Si$_2$ blocks
[Figs.~\ref{fig:band}(a),(b)]. First-principles calculations reveal a
spin-orbit coupling (SOC)-induced gap within an inverted Mo--Th band sector near the Fermi level
[Figs.~\ref{fig:band}(d),(e)]. Orbital-projected bands show that the bands forming the relevant
crossing along $\Gamma$--$M$ originate predominantly from Mo-$d_{xz/yz}$
and Th-$d_{x^2-y^2}$ orbitals (see Supplemental Material~\cite{SM}).
Without SOC, these states form a three-band crossing; SOC resolves it into
a protected Dirac crossing and a separate avoided crossing. The two
Kramers pairs at the Dirac point transform according to distinct
$C_4$ double-group representations,
so their hybridization is forbidden, whereas branches with the same
representations can gap. The orbital inversion across the latter gap is
accompanied by a Dirac-like feature in the calculated (001) surface
spectrum near $\bar\Gamma$ [Fig.~\ref{fig:band}(f)].

The metallic band structure lacks a global indirect insulating gap. Nevertheless, a direct gap separates the selected band manifold from higher-energy bands throughout the Brillouin zone, allowing its topology to be characterized~\cite{FuKaneParity,Po2017,HasanKane2010,QiZhang2011,Song2018}.
For the complete manifold below the relevant SOC gap, we obtain
$(\nu_0;\nu_1\nu_2\nu_3)=(0;111)$. These indices describe a normal-state
band subspace, rather than an insulating ground state or the topology
of the superconducting order. A minimal Mo--Th orbital model instead
gives $(1;000)$ for its corresponding low-energy gap. Including additional
C-$s,p_z$ orbitals with their hybridizations produces a model with
$(0;111)$ while preserving the low-energy surface feature~\cite{SM}. Despite their different $\mathbb{Z}_2$ indices, both models exhibit the Mo--Th-derived helical surface states found in the first-principles calculations.

\begin{figure}[t]
\centering
\includegraphics[width=1.0\columnwidth]{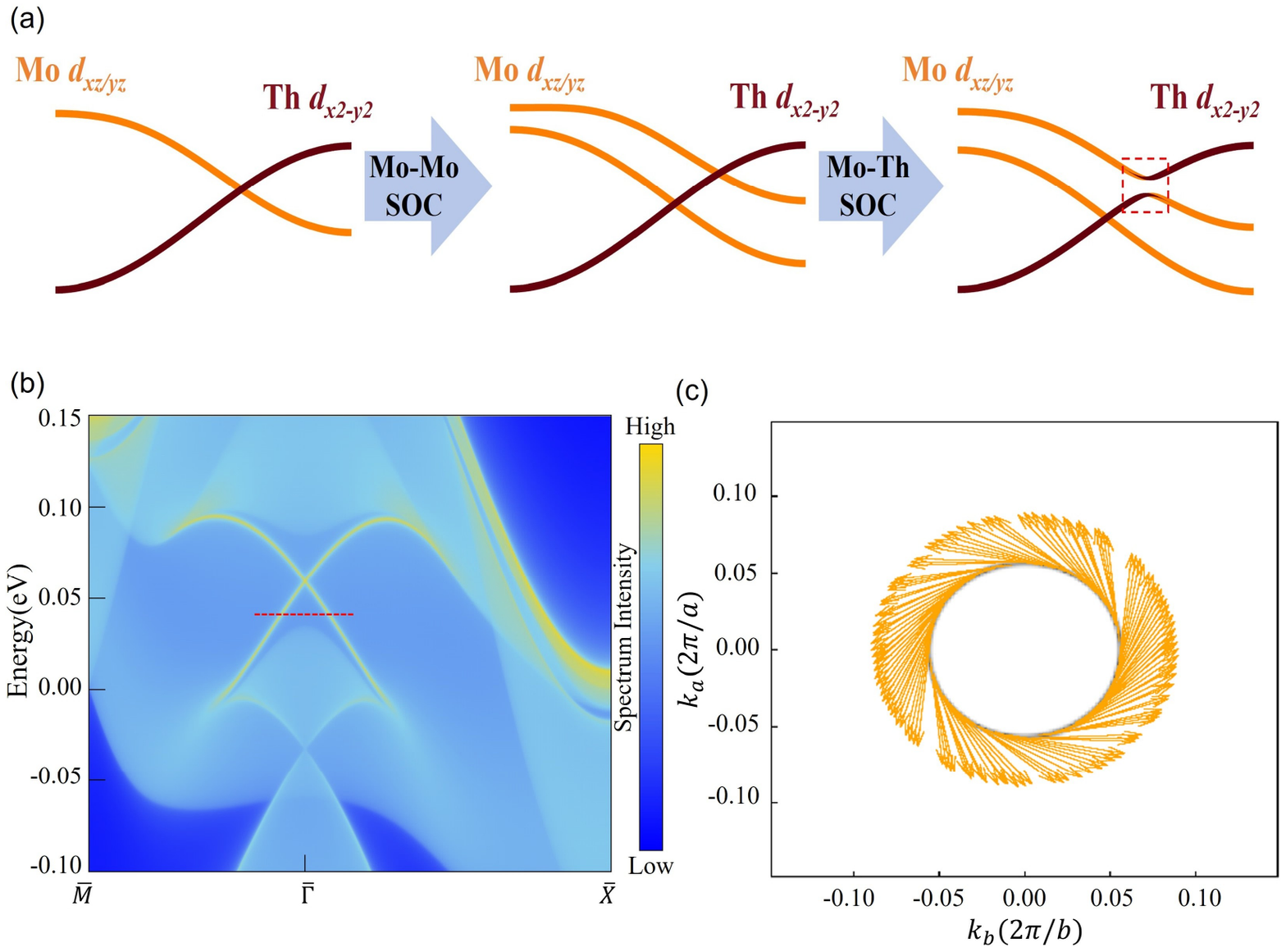}
\caption{SOC-induced band evolution and topological surface states obtained from the effective tight-binding Hamiltonian.
(a) Schematic of Mo-$d_{xz/yz}$ and Th-$d_{x^2-y^2}$ bands and the
roles of intra-Mo and interorbital Mo--Th SOC. (b) (001) surface
spectral function of the symmetry-based Mo--Th tight-binding model~\cite{SM}.
(c) In-plane spin texture at the energy indicated in (b).
}
\label{fig:mechanism}
\end{figure}

To identify the microscopic origin of the helical surface feature, we
compare SOC channels in a separate $24\times24$ downfolded
Wannier Hamiltonian. Removing the intra-Mo SOC block slightly reduces
the gap but leaves the Dirac-like (001) surface feature essentially
intact, whereas removing the Mo--Th SOC block eliminates this state~\cite{SM}. Guided by this independent test,
we construct the symmetry-based Mo-$d_{xz/yz}$/Th-$d_{x^2-y^2}$ model
illustrated in Fig.~\ref{fig:mechanism}(a). Its leading SOC terms have
the forms $H_{\mathrm{Mo}}^{\mathrm{SOC}}\sim\lambda_1s_zL_z$ and
$H_{\mathrm{Mo\text{-}Th}}^{\mathrm{SOC}}\sim
\lambda_3f(k) \mathcal{M}_3$ (See details in supplemental materials ~\cite{SM}). Thus, intra-Mo SOC
splits the Mo orbital branches without lifting the spin-Kramers degeneracy, whereas the Mo--Th term hybridizes the inverted
states and opens the relevant gap. The resulting surface band
[Fig.~\ref{fig:mechanism}(b)] carries a spin–momentum-locked in-plane spin texture
[Fig.~\ref{fig:mechanism}(c)], consistent with the termination-resolved
first-principles surface calculations~\cite{SM}. Together with the pre-existing bulk superconductivity, this
helical sector motivates an intrinsic, rather than heterostructure-based,
route to connate topological superconductivity and vortex Majorana modes.

\textit{Two microscopic routes to Kitaev vortex states.---}
Electron--phonon calculations~\cite{SM,AllenDynes1975} support a phonon mediated pairing channel. As a minimal assumption, we take an onsite spin-singlet s-wave order parameter. For a singly quantized vortex parallel to the crystallographic $c$ axis, we use
\begin{equation}
\Delta(\bm r;\bm r_0)=\Delta_0\tanh\!\left(
 \frac{|\bm r-\bm r_0|}{\xi}\right)
 e^{i\arg(\bm r-\bm r_0)},
\label{eq:vortex}
\end{equation}
where $\bm r_0$ specifies the vortex position relative to the atomic lattice. Translation along the vortex line permits a one-dimensional BdG description $\mathcal H_v(k_c;\bm r_0)$. Fixed vorticity breaks time reversal while preserving particle--hole symmetry, placing a generic fully gapped vortex line in class D~\cite{AltlandZirnbauer1997,Schnyder2008,Kitaev2009,TeoKane2010,LuoWu2025,WangWangHuangWuHao2026,WangLiHuangHao2024}. Its $\mathbb Z_2$ invariant is
\begin{equation}
\mathcal Q_v=\operatorname{sgn}\!\left[
\operatorname{Pf}B(0)\operatorname{Pf}B(\pi)\right],\quad
B(k_c)=\mathcal H_v(k_c)U_{\mathcal C},
\label{eq:pf}
\end{equation}
with the Pfaffians evaluated in a common
Majorana convention~\cite{Kitaev2001, Hosur2011}. The classification as a gapped Kitaev vortex also requires $E_{\rm vg}=\min_{k_c,n}|E_n(k_c)|>0$ throughout the one-dimensional Brillouin zone; in this case, the invariant
$\mathcal{Q}_v=-1$ ($+1$) identifies the
topological (trivial) phase. Endpoint Pfaffian signs alone do not characterize nodal or hybrid regimes~\cite{qin2019quasi,hu2022competing,LuoTian2026,LuoPanShiWu2025}. We focus on two gapped $\mathcal Q_v=-1$ regions with different microscopic origins, rather than on the previously established classification of gapless vortex phases.

\begin{figure}[t]
\centering
\includegraphics[width=1.0\columnwidth]{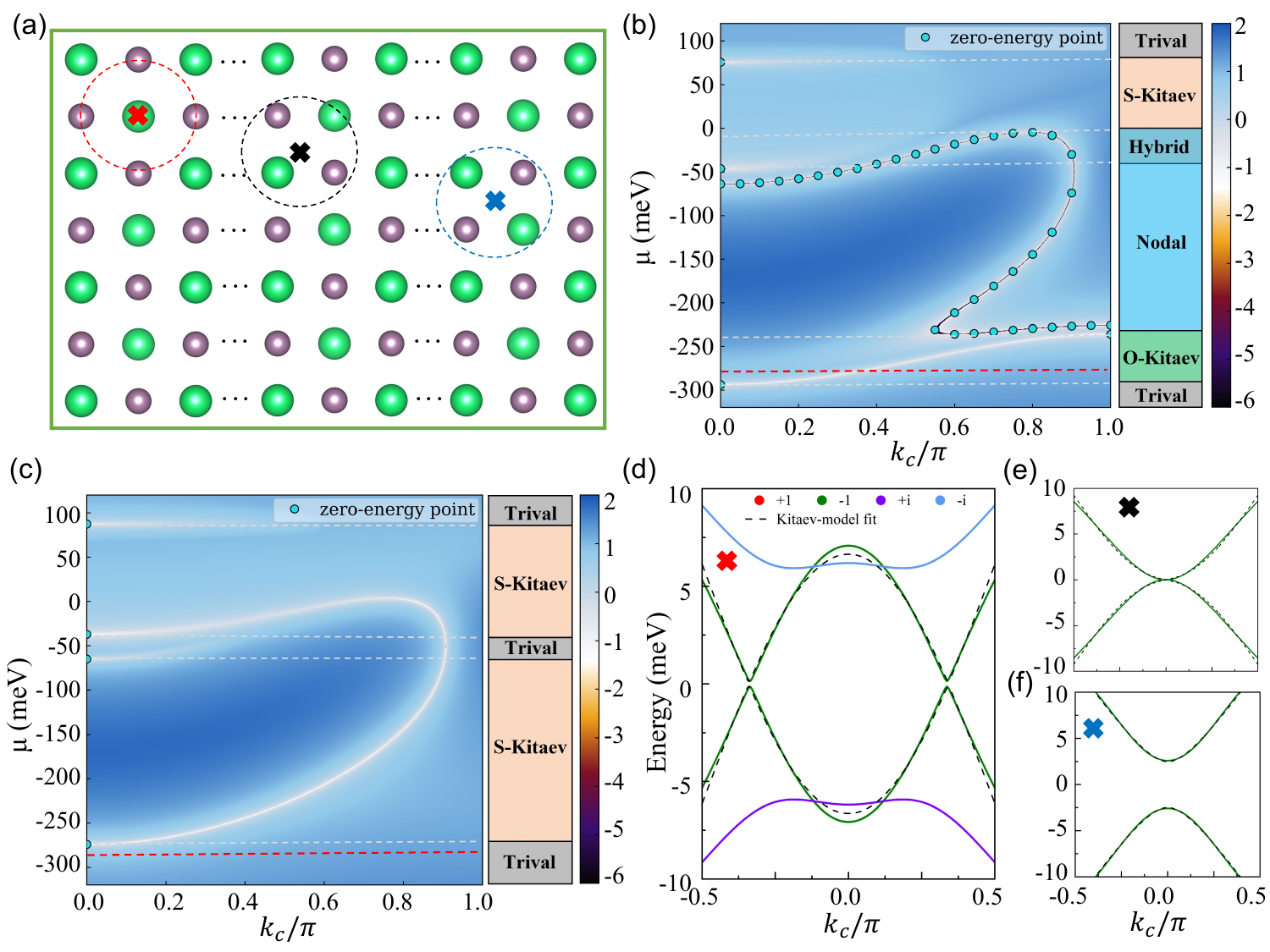}
\caption{Two Kitaev vortex regimes and
vortex-position-driven topological transition.
(a) Schematic of three independent
vortex-core configurations at inequivalent positions relative to
the atomic lattice, with Th (green) and Mo (purple) atoms.
Red, black, and blue crosses indicate the
Th-centered, intermediate, and displaced
vortex-core positions, respectively.
Dashed circles illustrate the corresponding
core regions.
(b),(c) Momentum-resolved vortex excitation-gap
maps versus chemical potential $\mu$ for the
Th-centered and displaced-core configurations,
respectively. The color scale represents the
logarithm of the minimum excitation energy,
and cyan circles mark zero-energy crossings.
S-Kitaev and O-Kitaev denote the surface-associated
and orbital-derived Kitaev regimes, respectively. The red dashed lines at $\mu=-280$~meV specify the spectra shown in (d) and (f).
(d)--(f) Evolution of the low-energy vortex BdG
spectrum across the position-driven topological
transition at $\mu=-280$~meV, corresponding to
the red, black, and blue core positions in (a).
In (d), the colored branches are resolved by the
combined $C_4$ eigenvalues, and black dashed
curves show the fitted effective Kitaev-chain
dispersion. Panels (e) and (f) illustrate the
gap closing and reopening during vortex displacement.
The phase diagrams are calculated using
$\Delta_0=100$~meV and $\xi=1.5a$.}
\label{fig:position}
\end{figure}

For a Th-centered vortex, the phase diagram reveals two separated Kitaev intervals [Fig.~\ref{fig:position}(b)]. The higher-chemical-potential interval, $-5\lesssim\mu\lesssim75.5$~meV, is associated with the inverted Mo--Th surface sector (S-Kitaev) and exhibits a relatively large vortex minigap. More unexpectedly, a distinct fully gapped region at $-294\lesssim\mu\lesssim-236$~meV has $\mathcal Q_v=-1$ despite being dominated by a Th-derived vortex-core orbital (O-Kitaev). The intervening nodal and hybrid regimes are gapless and are analyzed in the Supplemental Material~\cite{SM}. Thus S-Kitaev and O-Kitaev are not different symmetry classes, but two microscopic realizations of the same one-dimensional class-D topology.

The O-Kitaev mechanism begins with the local suppression of superconducting pairing. When the vortex is centered on a Th site, $\Delta(\bm r_{\rm Th})=0$, leaving a low-energy orbital-derived core level near $\pm|\varepsilon_{\rm Th}-\mu|$ before hybridization. Interlayer hopping and coupling to the surrounding paired orbitals generate a dispersive vortex-core band and an induced anomalous pairing. In an isolated effectively spinless channel, fermionic antisymmetry requires the projected pairing to be odd in $k_c$; the lowest harmonic is proportional to $\sin k_c$, although the original bulk pairing is even parity. The small induced pairing opens a narrow avoided-crossing gap and explains the much longer Majorana decay length of the O-Kitaev phase. The vanishing onsite order parameter alone is not a topological criterion: nontriviality is established by a fully gapped vortex-line spectrum and a nontrivial Pfaffian index.

To assess whether the orbital Kitaev phase requires helical surface states, we study a modified model in which the Mo-derived bands are shifted away from the Th bands, removing the original surface state. A c-axis Zeeman term separates the coincident vortex-band inversions, yielding a fully gapped $\mathcal Q_v=-1$ interval and end-localized zero modes (see Fig. S9 in the Supplemental Material~\cite{SM}). This result shows that the orbital Kitaev phase does not require the original surface Dirac cone, although Mo–Th hybridization and spin–orbit coupling remain present.

A direct test of this mechanism is provided by the response to vortex displacement. At fixed $\mu=-280$~meV, shifting the vortex center away from the Th site closes and reopens the vortex-line gap and reverses $\mathcal Q_v$ from $-1$ to $+1$ [Figs.~\ref{fig:position}(d)--(f)]. Along the same examined displacement path, the higher-energy surface-associated Kitaev interval remains nontrivial [Fig.~\ref{fig:position}(c)]. The two channels are therefore distinguished not only by their minigaps but also by their sensitivity to the vortex-core position.

The low-energy spectra in the O-Kitaev sector are reproduced by a fitted one-dimensional Kitaev Hamiltonian~\cite{Kitaev2001},
\begin{equation}
\begin{aligned}
H_{\rm core}^{\rm eff}(k_c;\bm r_0)={}&
\left[m(\bm r_0)+2t_z(\bm r_0)(1-\cos k_c)\right]\tau_z\\
&+\Delta_p(\bm r_0)\sin k_c\,\tau_y .
\end{aligned}
\label{eq:kitaev}
\end{equation}
For the fitted O-Kitaev, critical, and trivial spectra, the mass $m$ evolves through $-6.636$, $0$, and $2.490$~meV, respectively, while $m+4t_z$ remains positive. Thus $\mathcal Q_{\rm eff}=\operatorname{sgn}[m(m+4t_z)]$ switches at a $k_c=0$ mass closing, in agreement with the effective-model description of the full-BdG transition. The corresponding fitted odd-momentum pairing increases from $\Delta_p=0.153$ to $1.457$ and $5.499$~meV. For a displacement $d=|\bm r_{\rm Th}-\bm r_0|$, the local order parameter recovers as $|\Delta_{\rm Th}|=\Delta_0\tanh(d/\xi)$, reconstructing the effective channel through its local pairing and hybridization. 

\begin{figure}[t]
\centering
\includegraphics[width=1.0\columnwidth]{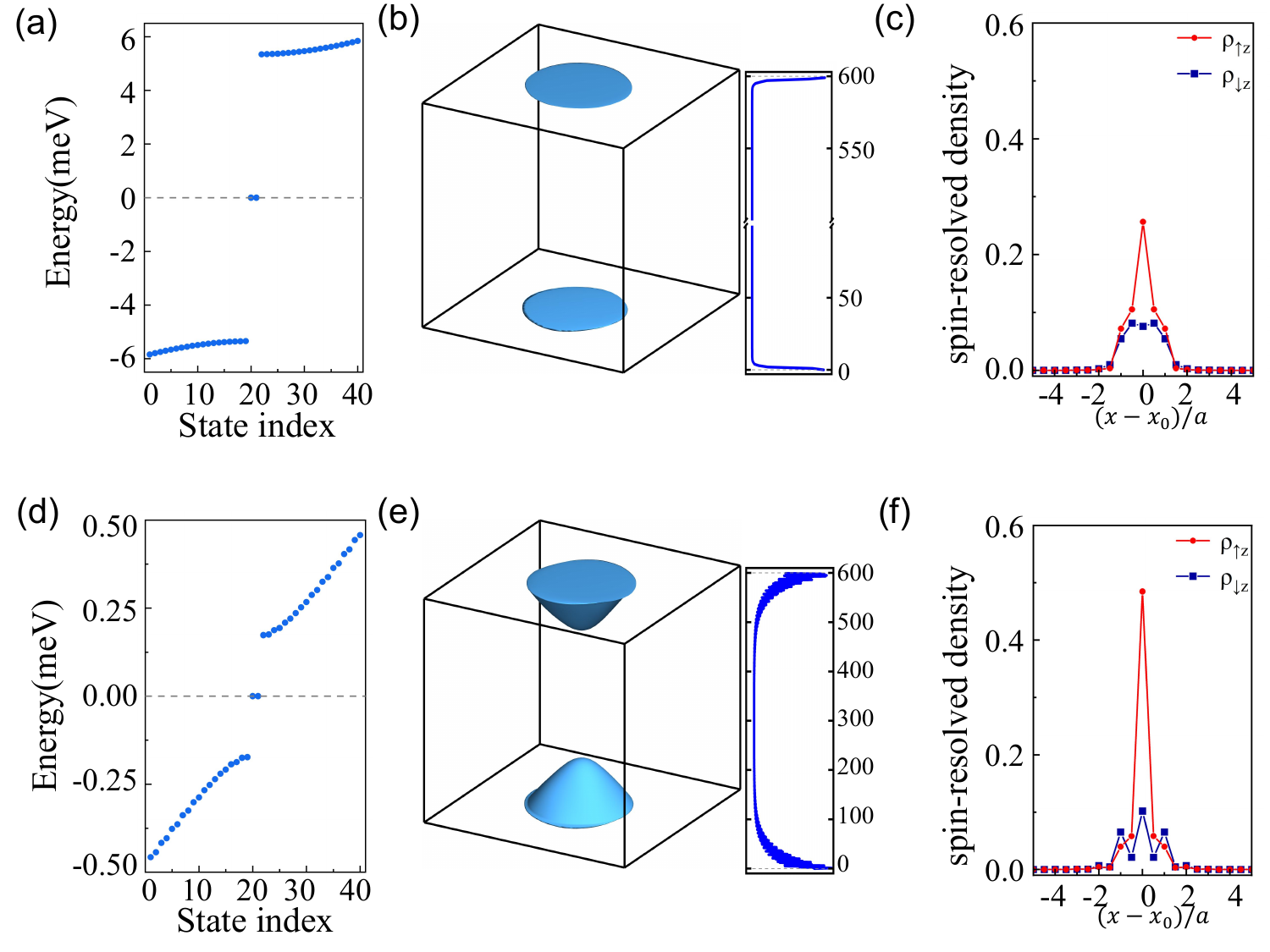}
\caption{Majorana end modes in two
microscopically distinct Kitaev vortex phases.
(a)--(c) S-Kitaev
phase at $\mu=60$~meV:
(a) low-energy eigenvalues of the finite-length
vortex BdG Hamiltonian;
(b) three-dimensional probability distribution
of the Majorana end modes, together with the
layer-resolved density along the vortex line;
(c) near-surface spin-resolved electron densities
$\rho_{\uparrow z}$ and $\rho_{\downarrow z}$
along the $x$ direction, summed over the
outermost two atomic layers.
(d)--(f) Corresponding results for the
O-Kitaev phase
at $\mu=-260$~meV.
Both phases support near-zero-energy Majorana
modes localized at opposite ends of the vortex
line, but exhibit markedly different excitation
gaps, longitudinal localization lengths, and
spin-resolved spatial profiles.
The finite-vortex calculations use
$15\times15\times600$ lattice sites.}
\label{fig:finite}
\end{figure}

\textit{Majorana signatures and broader implications.---}
Open boundaries along $c$ produce near-zero-energy modes localized at the two vortex terminations in both Kitaev regions [Fig.~\ref{fig:finite}]. The S-Kitaev end states are more tightly confined along the vortex, whereas the small induced O-Kitaev gap leads to more extended Majorana wave functions. Their near-surface spin-projected electron weights also differ, providing complementary spectroscopic signatures [Figs.~\ref{fig:finite}(c),(f)]. In particular, the calculated out-of-plane spin selectivity and distinct lateral profiles motivate spin-polarized scanning tunneling microscopy and spectroscopy (STM/STS) studies of vortices pinned at inequivalent atomic positions~\cite{PhysRevLett.116.257003,WangWiebeGu2021,PanLuoGaoLiu2022,JackXieYazdani2021}. Two well-separated vortices with distinct local core positions could realize different topological sectors without experimentally translating a single vortex through the entire switching path~\cite{LiuFranz2015,Biswas2013,LiuWanYang2024}. 

The weak O-Kitaev minigap is an important limitation of the present material realization. Our parameter scans demonstrate that characteristic low-energy inversions survive over a range of enhanced pairing amplitudes and core sizes (See Supplemental Material~\cite{SM}), but do not establish a fully gapped adiabatic path to the experimentally estimated $\Delta$ and $\xi$. At these scales, the small induced gap and the large system size needed for calculations make it difficult to predict whether the zero modes can be observed experimentally. Our calculations establish the position-driven
topological transition in the material-based model, while
its realization at experimental parameters remains to be tested. This points toward materials with stronger effective spin-orbit coupling and favorable interorbital or interlayer hybridization, which may enlarge the induced orbital-channel pairing gap (See Supplemental Material~\cite{SM}). 
The control calculation in Fig. S9 of the Supplemental Material~\cite{SM} shows that the orbital-derived route does not require the original helical surface state, suggesting opportunities beyond established topological-surface
platforms.

\textit{Conclusion.---}
We identify the synthesized intergrowth superconductor Th$_2$Mo$_2$Ir$_2$Si$_4$C as a candidate connate topological platform and uncover two microscopic routes to Majorana vortex end modes: the conventional inverted-surface-state route and an additional orbital-derived core route. The latter realizes an effective Kitaev channel whose topology is switched by vortex positioning through a low-energy mass inversion. A surface-state-free control
calculation further supports its distinct microscopic origin. While its experimental observability in this compound
remains open, the orbital-derived route and its position--driven topological transition suggest a way to design Majorana vortices in other multiorbital superconductors.

\begin{acknowledgments}
We thank Hai-Hu Wen and Guang-Han Cao for helpful discussions. This work was financially supported by the National Key R\&D Program of China (Grants No. 2022YFA1403200 and 2024YFA1613200), the National Natural Science Foundation of China (Grants No. 92265104, 12022413 and 12604254), the Basic Research Program of the Chinese Academy of Sciences Based on Major Scientific Infrastructures (Grant No. JZHKYPT-2021-08), the CASHIPS Directors Fund (Grant No. BJPY2023A09), Anhui Provincial Major S\&T Project (Grant No. s202305a12020005), and the High Magnetic Field Laboratory of Anhui Province (Grant No. AHHM-FX-2020-02).
\end{acknowledgments}

%


\clearpage
\onecolumngrid
\clearpage
\setcounter{equation}{0}
\renewcommand{\theequation}{S\arabic{equation}}
\setcounter{figure}{0}
\renewcommand{\thefigure}{S\arabic{figure}}
\setcounter{table}{0}
\renewcommand{\thetable}{S\Roman{table}}
\renewcommand{\theHfigure}{SMfig.\arabic{figure}}
\renewcommand{\theHequation}{SMeq.\arabic{equation}}
\renewcommand{\theHtable}{SMtab.\arabic{table}}

\let\ThMoOriginalIncludeGraphics\includegraphics
\renewcommand{\includegraphics}[2][]{%
  \IfFileExists{#2}{\ThMoOriginalIncludeGraphics[#1]{#2}}{%
  \fbox{\parbox[c][32mm][c]{0.74\linewidth}{\centering\small
       Original graphic file not supplied:\par\texttt{\detokenize{#2}}}}%
  }%
}

\begin{center}
{\bfseries\large Supplemental Material for ``\REV{Connate Topological Superconductivity and Vortex-Position-Controlled Majorana Modes in the Intergrowth Superconductor Th$_2$Mo$_2$Ir$_2$Si$_4$C}''}\par
\vspace{7pt}
\end{center}
\vspace{10pt}

\section{S1. First-Principles Electronic-Structure Calculations}

The first-principles electronic-structure calculations were performed within density functional theory (DFT) using the Vienna \textit{Ab initio} Simulation Package (VASP)~\cite{SM:KresseFurthmuller1996}. The exchange--correlation energy was treated within the generalized-gradient approximation (GGA) in the Perdew--Burke--Ernzerhof (PBE) form~\cite{SM:PBE1996}, and the interaction between valence electrons and ionic cores was described using the projector augmented-wave (PAW) method~\cite{SM:Blochl1994,SM:KresseJoubert1999}. Starting from the experimentally refined crystal structure of Th$_2$Mo$_2$Ir$_2$Si$_4$C~\cite{SM:LiJACS2025}, both the lattice parameters and internal atomic coordinates were fully relaxed until the residual Hellmann--Feynman force on each atom was smaller than $1\times10^{-3}$~eV/\AA\ and the total-energy change between successive ionic steps was below $1\times10^{-7}$~eV. The relaxed structure was subsequently used for all calculations discussed below. Brillouin-zone integrations were carried out using a $\Gamma$-centered Monkhorst--Pack mesh~\cite{SM:MonkhorstPack1976}; a $13\times13\times13$ mesh was used for the bulk self-consistent calculations, together with a plane-wave kinetic-energy cutoff of 550~eV. Calculations both without and with spin--orbit coupling (SOC) were performed as required for the analysis of the band crossings and SOC-induced topological gap.

For the cleavage-energy calculations, slabs were constructed by stacking two crystallographic unit cells along the [001] direction and separating periodically repeated slabs by a vacuum region of 25~\AA. A denser $15\times15\times1$ $\Gamma$-centered $k$-point mesh was used for the two-dimensional Brillouin-zone sampling, and the electronic self-consistency threshold was tightened to $1\times10^{-8}$~eV. Because the two exposed surfaces of the cleavage slabs are generally inequivalent, a dipole correction was applied along the crystallographic $c$ direction to suppress spurious electrostatic interactions between periodically repeated asymmetric slabs~\cite{SM:NeugebauerScheffler1992}.

\subsection{Orbital-projected band structure and \textsc{Wannier} functions}

\begin{figure}[htbp]
	\centering
	\includegraphics[width=0.8\linewidth]{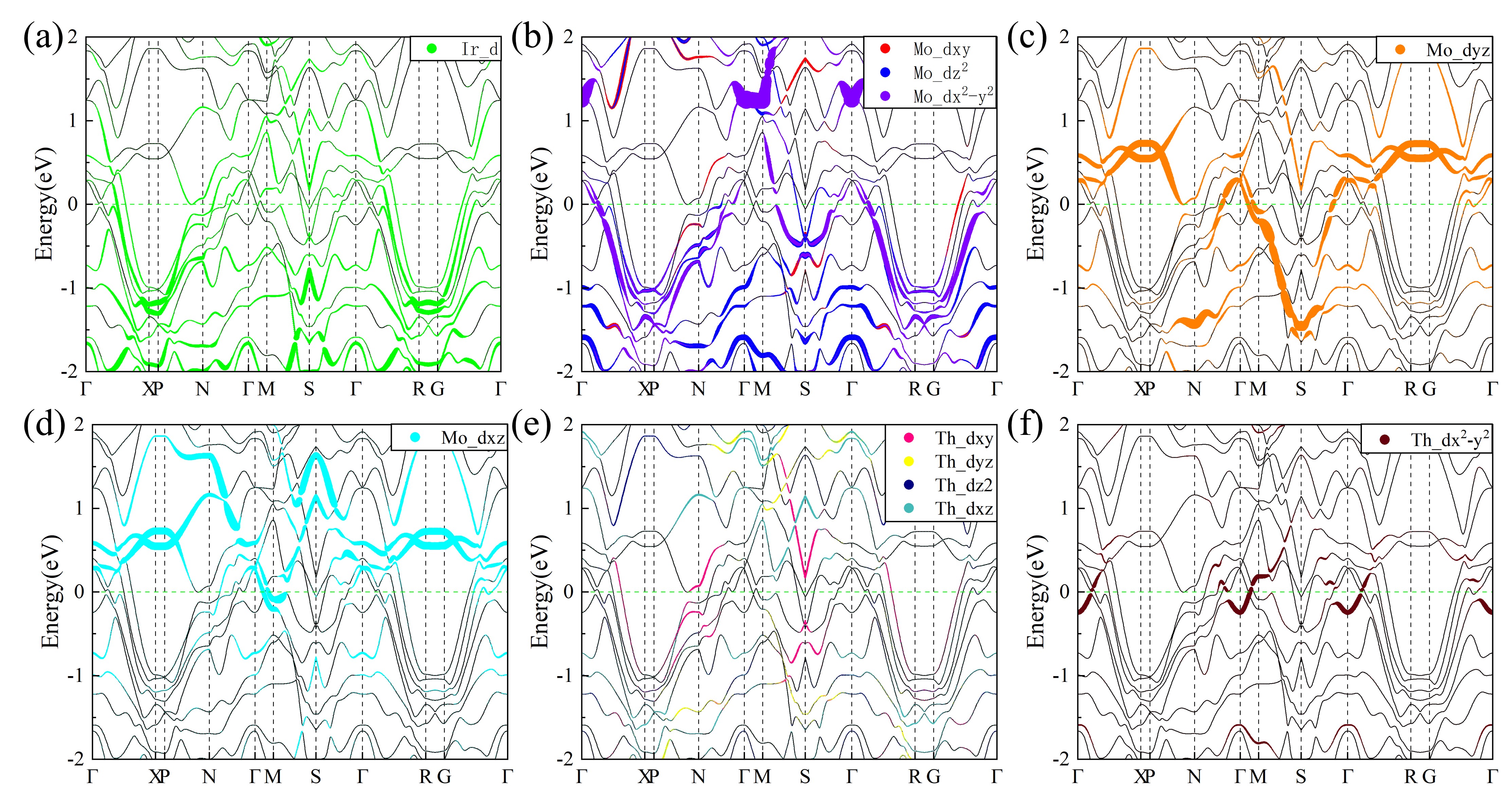}
	\caption{\REV{Orbital-projected band structures of Th$_2$Mo$_2$Ir$_2$Si$_4$C including SOC. Panels (a)--(f) display element- and orbital-resolved spectral weights of the Ir, Mo, and Th $d$ states, highlighting the Mo-$d_{xz/yz}$ and Th-$d_{x^2-y^2}$ contributions to the inverted low-energy bands.}}
	\label{fig:S_projected_bands}
\end{figure}

To construct an accurate low-energy tight-binding representation, maximally localized Wannier functions (MLWFs)~\cite{SM:MarzariVanderbilt1997} were generated using the \textsc{Wannier90} package~\cite{SM:Mostofi2008,SM:Mostofi2014}. The initial projection manifold included the Mo-$d$, Th-$d$, and Ir-$d$ orbitals, together with the C-$p$ and Si-$p$ orbitals, in order to faithfully reproduce the first-principles bands in the energy window surrounding the Fermi level. The (001) surface spectral functions and momentum-resolved spin textures were evaluated from the Wannier-based tight-binding Hamiltonian using \textsc{WannierTools}~\cite{SM:WuWannierTools2018}.

\subsection{Spin--orbit-coupling origin of the (001) topological surface states}

To identify the microscopic SOC channel responsible for the topological surface states and thereby guide the construction of the low-energy effective model, we downfolded the full MLWF Hamiltonian onto the spinful Mo-$d_{xz/yz}$ and Th-$d_{x^{2}-y^{2}}$ target subspace.  Let $P$ project onto these retained orbitals and $Q=1-P$ onto the eliminated Wannier orbitals.  In the corresponding block representation,
\begin{equation}
	H(\mathbf{k})=
	\begin{pmatrix}
		H_{PP}(\mathbf{k}) & H_{PQ}(\mathbf{k})\\
		H_{QP}(\mathbf{k}) & H_{QQ}(\mathbf{k})
	\end{pmatrix}.
	\label{eq:S_lowdin_blocks}
\end{equation}
The downfolding process gives the $24\times24$ effective Hamiltonian
\begin{equation}
	H_{\mathrm{eff}}(\mathbf{k};E_{\mathrm{ref}})
	=
	H_{PP}(\mathbf{k})
	H_{PQ}(\mathbf{k})
	\bigl[E_{\mathrm{ref}}-H_{QQ}(\mathbf{k})\bigr]^{-1}
	H_{QP}(\mathbf{k}),
	\label{eq:S_lowdin_downfolding}
\end{equation}
where $E_{\mathrm{ref}}$ is chosen in the vicinity of the target topological gap.  The second term retains, to the downfolding order, the virtual processes through the eliminated Wannier sector and therefore incorporates their renormalization of the Mo--Th low-energy manifold.

\begin{figure}[htbp]
	\centering
	\includegraphics[width=0.6\linewidth]{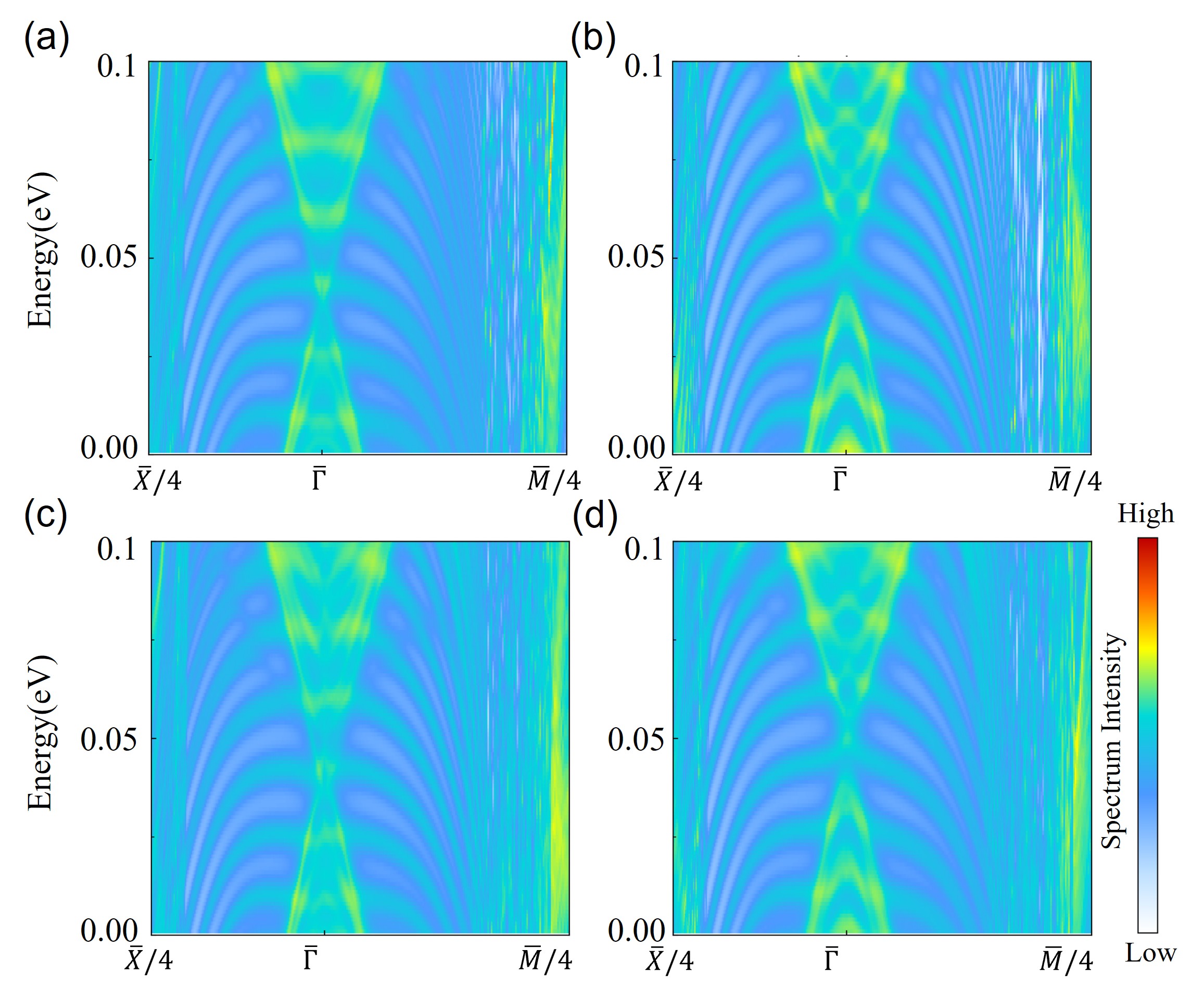}
	\caption{SOC-resolved (001) surface spectral functions obtained from the $24\times24$ downfolded Hamiltonian.  (a) Full SOC retained.  (b) All SOC matrix blocks removed.  (c) Mo--Mo SOC block removed while the remaining SOC terms are retained.  (d) Mo--Th coupling SOC block removed while the remaining SOC terms are retained.}
	\label{fig:S_SOC_surface_analysis}
\end{figure}

\section{S2. Parity Analysis and Symmetry Indicators}

The topological indices were evaluated for the band manifold below the target gap between the 88th and 89th bands. Accordingly, bands 1--88 were treated as occupied, while the next Kramers pair was excluded. We determine the topology independently using the Fu--Kane parity criterion and the symmetry indicators of space group $I4/mmm$ (No.~139).

\subsection{Fu--Kane parity analysis}

Because Th$_2$Mo$_2$Ir$_2$Si$_4$C crystallizes in the centrosymmetric space group $I4/mmm$ (No.~139) and preserves time-reversal symmetry in the normal state, the strong and weak $\mathbb Z_2$ indices can be obtained from the inversion eigenvalues of the occupied Kramers pairs using the Fu--Kane parity criterion~\cite{SM:FuKaneParity}. At each time-reversal-invariant momentum (TRIM) $\Lambda_i$, we define
\begin{equation}
\delta_i=\prod_{m=1}^{N_{\rm occ}/2}\xi_m(\Lambda_i),
\label{eq:S2_delta}
\end{equation}
where $\xi_m(\Lambda_i)=\pm1$ is the parity eigenvalue of one state in the $m$th occupied Kramers pair. The strong invariant is
\begin{equation}
(-1)^{\nu_0}=\prod_{i=1}^{8}\delta_i,
\label{eq:S2_nu0}
\end{equation}
and the three weak invariants satisfy
\begin{equation}
(-1)^{\nu_j}=
\prod_{\substack{n_j=1\\ n_{l\neq j}=0,1}}
\delta_{n_1n_2n_3},
\qquad j=1,2,3.
\label{eq:S2_nuj}
\end{equation}

In the primitive reciprocal-coordinate convention used in the VASP2Trace calculation, the eight TRIMs of the body-centered tetragonal Brillouin zone form four symmetry-equivalent sets, represented by $\Gamma$, $M$, $X$, and $N$. The calculated numbers of even- and odd-parity occupied Kramers pairs up to band 88 are summarized in Table~\ref{tab:S2_parity}.

\begin{table}[h]
\centering
\caption{Parity analysis for the occupied manifold consisting of bands 1--88. The coordinates are given in units of the primitive reciprocal lattice vectors. The multiplicity denotes the number of symmetry-equivalent TRIMs in the full Brillouin zone.}
\begin{tabular}{cccccc}
\toprule
TRIM & Representative $\mathbf{k}$ & Multiplicity & $n_+$ & $n_-$ & $\delta=(-1)^{n_-}$ \\
\midrule
$\Gamma$ & $(0,0,0)$ & 1 & 23 & 21 & $-1$ \\
$M$ & $(\tfrac12,\tfrac12,\tfrac12)$ & 1 & 21 & 23 & $-1$ \\
$X$ & $(0,0,\tfrac12)$ & 2 & 22 & 22 & $+1$ \\
$N$ & $(0,\tfrac12,0)$ & 4 & 22 & 22 & $+1$ \\
\bottomrule
\end{tabular}
\label{tab:S2_parity}
\end{table}

Substituting the parity products in Table~\ref{tab:S2_parity} into Eqs.~\eqref{eq:S2_nu0} and \eqref{eq:S2_nuj} gives
\begin{equation}
(\nu_0;\nu_1\nu_2\nu_3)=(0;111).
\label{eq:S2_final}
\end{equation}

\subsection{Symmetry-indicator analysis}

As an independent check, the same band manifold below the gap between the 88th and 89th bands was analyzed within the symmetry-indicator framework~\cite{SM:Po2017,SM:Song2018}. For space group $I4/mmm$ (No.~139), the independent symmetry-indicator group is $\mathbb Z_2\times\mathbb Z_8$. For direct comparison with the Fu--Kane result, we evaluate the three weak indicators $z_{2w,j}$ together with the inversion indicator $z_4$ and the tetragonal indicator $z_8$.

The weak indices are obtained from the odd-parity Kramers-pair counts,
\begin{equation}
z_{2w,j}
=
\sum_{\substack{K\in{\rm TRIM}\\ k_j=\pi}}
n^-_K
\pmod 2,
\qquad j=1,2,3,
\end{equation}
while the inversion indicator is
\begin{equation}
z_4=
\sum_{K\in{\rm TRIM}}
\frac{n^-_K-n^+_K}{2}
\pmod 4 .
\end{equation}
The additional tetragonal indicator is evaluated from the double-valued irreducible representations according to~\cite{SM:Song2018},
\begin{equation}
z_8=
\frac{3}{2}n^+_{3/2}
-\frac{3}{2}n^-_{3/2}
-\frac{1}{2}n^+_{1/2}
+\frac{1}{2}n^-_{1/2}
\pmod 8 .
\end{equation}
Using the parity counts in Table~\ref{tab:S2_parity} and the double-valued representation multiplicities of bands 1--88 gives
\begin{equation}
(z_{2w,1},z_{2w,2},z_{2w,3};z_4,z_8)
=
(1,1,1;0,0).
\end{equation}
This result is fully consistent with the Fu--Kane result in Eq.~\eqref{eq:S2_final}; the two analyses therefore independently identify the same weak-topological phase.

\section{S3. Cleavage Planes, Termination-Resolved Surface Spectra, and Cleavage Energetics}

The main text presents a representative (001) surface spectrum. In this section, we provide a more complete description of the possible (001) cleavage planes, the corresponding top and bottom terminations, and their termination-resolved surface spectra. We also compare their cleavage energies to assess the relative energetic accessibility of these surfaces.

\subsection{Candidate (001) cleavage planes and surface terminations}

\begin{figure}[htbp]
	\centering
	\includegraphics[width=0.6\linewidth]{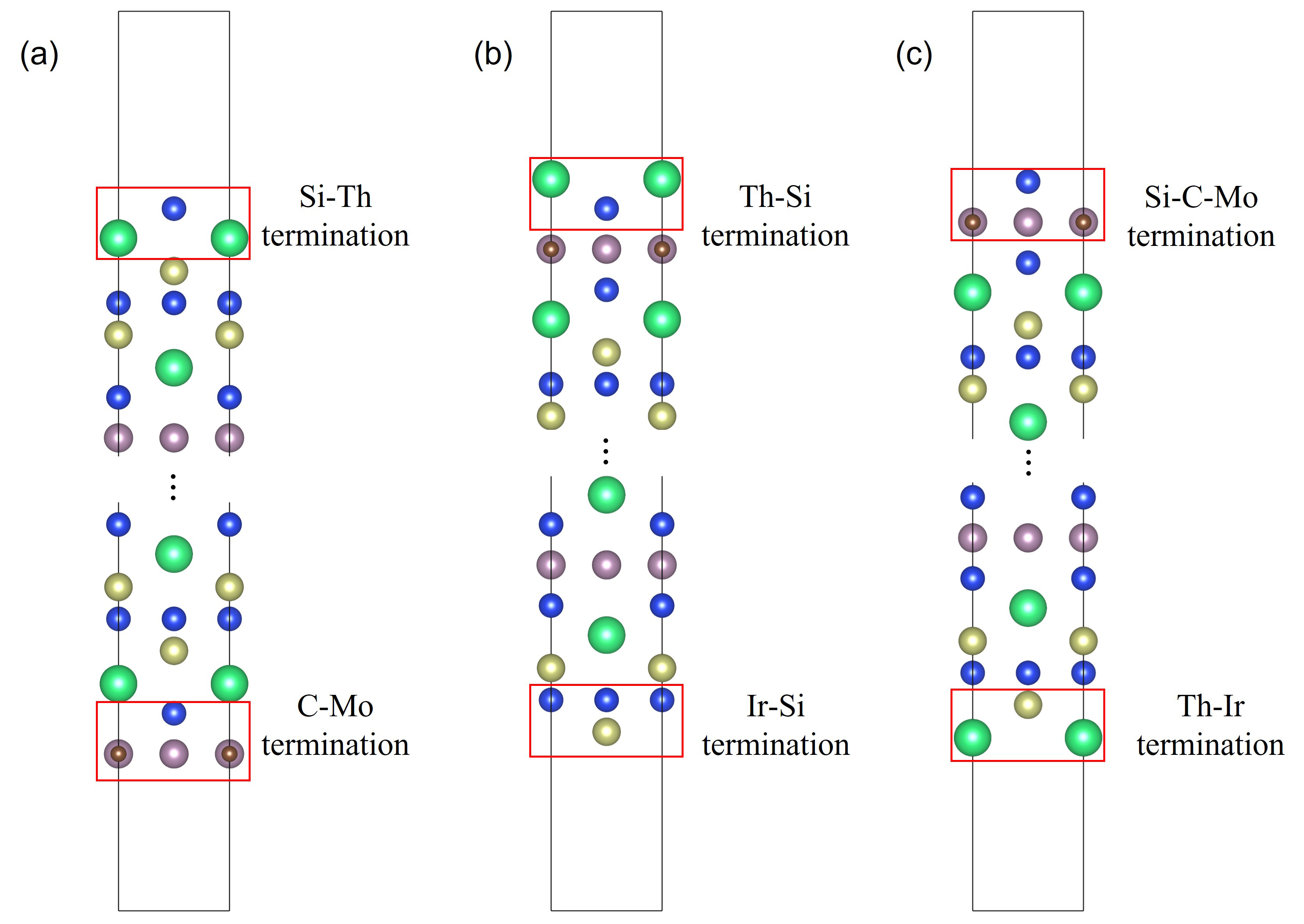}
	\caption{
		Schematic illustration of the three candidate (001) cleavage planes of Th$_2$Mo$_2$Ir$_2$Si$_4$C. 
		Panels (a)--(c) correspond to cleavage plane 1, cleavage plane 2, and cleavage plane 3, respectively. 
		Each cleavage generates two complementary terminations on the top and bottom surfaces, giving six termination-resolved surface configurations in total.
	}
	\label{fig:S_cleavage_planes}
\end{figure}

\subsection{Termination-resolved surface spectra}

\begin{figure}[h]
	\centering
	\includegraphics[width=0.95\linewidth]{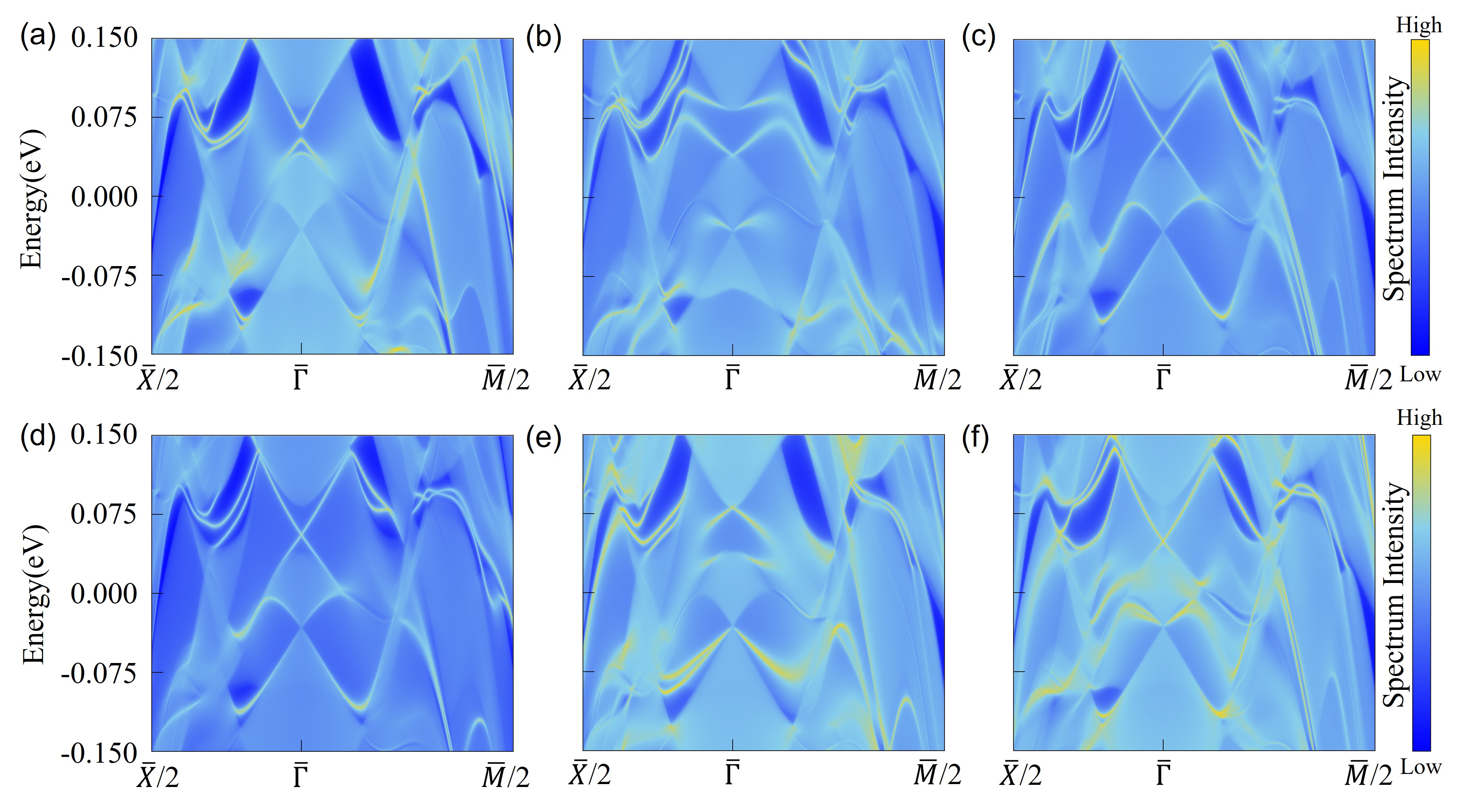}
	\caption{
		Termination-resolved (001) surface spectra for the three cleavage planes.
		Panels (a) and (d) correspond to cleavage plane~1, showing the Si--Th and C--Mo terminations, respectively.
		Panels (b) and (e) correspond to cleavage plane~2, showing the Th--Si and Ir--Si terminations, respectively.
		Panels (c) and (f) correspond to cleavage plane~3, showing the Si--C--Mo and Th--Ir terminations, respectively.
		The C--Mo termination in panel (d) is the representative surface shown in the main text.
		The Si--C--Mo and Th--Ir terminations in panels (c) and (f) also exhibit Dirac-like linearly dispersing surface states.
	}
	\label{fig:S_surface_terminations}
\end{figure}

The surface spectra for the six terminations generated by the three cleavage planes are summarized in Fig.~\ref{fig:S_surface_terminations}. The representative spectrum shown in the main text corresponds to the bottom surface of cleavage plane~1, namely the C--Mo termination. In addition to this representative termination, Dirac-like linearly dispersing surface states are also found for the Si--C--Mo and Th--Ir terminations associated with cleavage plane~3. These results show that the topological surface connectivity is not restricted to a single termination, although the detailed spectral weight and energy position depend on the surface termination.

\subsection{Cleavage energies}

To compare the relative energetic accessibility of the three cleavage configurations, we calculated their cleavage energies using stoichiometric slabs. For a slab generated by separating the bulk into two complementary surfaces, the work of separation is defined as
\begin{equation}
	E_{\rm cleav} =
	\frac{E_{\rm slab}-N E_{\rm bulk}}{A}
	=
	\gamma_{\rm top}+\gamma_{\rm bottom},
	\label{eq:S_cleavage_energy}
\end{equation}
where $E_{\rm slab}$ is the total energy of the cleavage slab, $E_{\rm bulk}$ is the energy of the corresponding bulk unit, $N$ is the number of bulk units contained in the slab, $A$ is the surface area, and $\gamma_{\rm top}$ and $\gamma_{\rm bottom}$ are the surface energies of the two complementary terminations~\cite{SM:WangNiAlCleavage2020}. This definition is the work needed to create the two surfaces together. For a symmetric slab with identical top and bottom surfaces, it reduces to the commonly used expression $\gamma=(E_{\rm slab}-N E_{\rm bulk})/(2A)$~\cite{SM:Sun2013SurfaceSlabs,SM:Tran2016SurfaceEnergies}.

\begin{table}[h]
	\centering
	\caption{Calculated cleavage energies of the three candidate (001) cleavage planes.}
	\begin{tabular}{cc}
		\toprule
		Cleavage plane & $E_{\rm cleav}$ (eV/\AA$^2$)\\
		\midrule
		1 & 0.2817\\
		2 & 0.2779\\
		3 & 0.2809\\
		\bottomrule
	\end{tabular}
	\label{tab:S_cleavage}
\end{table}

The three cleavage energies are very close. The lowest value is obtained for cleavage plane~2, whereas cleavage planes~1 and 3 are higher by only $3.8\times10^{-3}$ and $3.0\times10^{-3}$~eV/\AA$^2$, respectively. These differences correspond to approximately $1.4\%$ and $1.1\%$ of the cleavage energy scale. Therefore, within the accuracy of the present slab calculations, the three candidate cleavages should be regarded as energetically competitive rather than sharply separated. This near degeneracy suggests that more than one termination may be experimentally accessible, depending on details of sample preparation, fracture conditions, and surface relaxation.

\section{S4. Low-Energy Effective Tight-Binding Model}

The superconducting and vortex-line calculations in the main text are based on a low-energy effective tight-binding model constructed in the primitive-cell representation. The model is designed to capture the Mo--Th band inversion, the SOC-induced target gap, and the surface states relevant to the BdG calculations.

\subsection{Primitive-cell basis and spinless Hamiltonian}

We retain the Mo $d_{xz}/d_{yz}$ orbitals and the Th $d_{x^2-y^2}$ orbital, which dominate the inverted low-energy bands discussed in the main text. In the primitive cell, the spinless basis is chosen as
\begin{equation}
	\Phi_0 =
	\left(
	|{\rm Mo1},d_{xz}\rangle,
	|{\rm Mo1},d_{yz}\rangle,
	|{\rm Mo2},d_{xz}\rangle,
	|{\rm Mo2},d_{yz}\rangle,
	|{\rm Th1},d_{x^2-y^2}\rangle,
	|{\rm Th2},d_{x^2-y^2}\rangle
	\right)^T .
	\label{eq:S_primitive_basis_spinless}
\end{equation}
Including spin, the full basis is
\begin{equation}
	\Phi =
	(\Phi_{0,\uparrow},\Phi_{0,\downarrow})^T ,
	\label{eq:S_primitive_basis_spinful}
\end{equation}
which gives a 12-band spinful Hamiltonian in the primitive-cell representation.

The spin-independent part of the Hamiltonian is written as
\begin{equation}
	H_0(\mathbf{k})=s_0\otimes h_0(\mathbf{k}),
	\label{eq:S_H0_spinful}
\end{equation}
where $\mathbf{k}=(k_1,k_2,k_3)$ denotes the primitive reciprocal coordinates. The $6\times6$ spinless Hamiltonian is
\begin{equation}
	h_0(\mathbf{k})=
	\begin{pmatrix}
		\epsilon_1 & 0 & m & 0 & u_1 & u_2 \\
		0 & \epsilon_2 & 0 & m & 0 & 0 \\
		m & 0 & \epsilon_2 & 0 & 0 & 0 \\
		0 & m & 0 & \epsilon_1 & v_1 & v_2 \\
		u_1^{*} & 0 & 0 & v_1^{*} & \epsilon_3 & \tau \\
		u_2^{*} & 0 & 0 & v_2^{*} & \tau^{*} & \epsilon_3
	\end{pmatrix}.
	\label{eq:S_h0_primitive}
\end{equation}
The matrix elements are defined as
\begin{align}
	\epsilon_1 &=
	e_1+2t_{N11}\left[\cos(k_1+k_3)+\cos(k_2+k_3)\right],
	\\
	\epsilon_2 &=
	e_2+2t_{N22}\left[\cos(k_1+k_3)+\cos(k_2+k_3)\right],
	\\
	\epsilon_3 &=
	e_3+2t_{N33}\left[\cos(k_1+k_3)+\cos(k_2+k_3)\right]
	\nonumber\\
	&\quad
	-2t_{c37}\left[
	\cos k_1+\cos k_2+\cos k_3+\cos(k_1+k_2+k_3)
	\right],
	\\
	m &=
	-2t_{c11}\cos\frac{k_1+k_2}{2}
	+4t_{N121}\cos\frac{k_1+k_3}{2}\cos\frac{k_2+k_3}{2},
	\\
	u_1 &=
	-2i t_{N15}e^{i\phi}
	\sin\frac{k_2+k_3}{2},
	\qquad
	u_2 =
	2i t_{N15}e^{-i\phi}
	\sin\frac{k_2+k_3}{2},
	\\
	v_1 &=
	2i t_{N15}e^{i\phi}
	\sin\frac{k_1+k_3}{2},
	\qquad
	v_2 =
	-2i t_{N15}e^{-i\phi}
	\sin\frac{k_1+k_3}{2},
	\\
	\tau &= t_{N56}e^{-2i\phi},
	\qquad
	\phi=\frac{13}{100}(k_1+k_2).
\end{align}
The parameter $t_{N15}$ describes the Mo--Th hybridization in the retained low-energy subspace, while $t_{c11}$ and $t_{c37}$ encode inter-unit hopping processes that are important for reproducing the dispersion of the inverted bands.

\subsection{Spin--orbit coupling terms and symmetry constraints}

The spinful effective Hamiltonian is written as
\begin{equation}
	H_{\rm eff}(\mathbf{k})
	=
	H_0(\mathbf{k})
	+
	H_{\rm SOC}(\mathbf{k}),
	\qquad
	H_{\rm SOC}(\mathbf{k})=H_{\lambda_1}+H_{\lambda_3}(\mathbf{k}),
	\label{eq:S_Heff_primitive}
\end{equation}
where $\mathbf{k}=(k_1,k_2,k_3)$ denotes the primitive reciprocal coordinates.

The first SOC term is the onsite atomic SOC between the Mo $d_{xz}$ and $d_{yz}$ orbitals. In the primitive-cell basis, it is written as
\begin{equation}
	H_{\lambda_1}
	=
	\lambda_1 s_z\otimes L_{\rm Mo},
	\label{eq:S_Hlambda1}
\end{equation}
where
\begin{equation}
	L_{\rm Mo}
	=
	i\left(
	|1\rangle\langle2|-|2\rangle\langle1|
	+
	|3\rangle\langle4|-|4\rangle\langle3|
	\right).
	\label{eq:S_LMo}
\end{equation}
Here $|1\rangle,\ldots,|6\rangle$ denote the spinless orbitals defined in Eq.~\eqref{eq:S_primitive_basis_spinless}. Thus, $H_{\lambda_1}$ acts only within the Mo $d_{xz}/d_{yz}$ subspace and is absent on the Th orbitals.

The interorbital SOC was generated directly in the primitive-cell representation by imposing the full space group $I4/mmm$ (No.~139).  The matrix that converts primitive fractional coordinates to the conventional crystallographic frame is
\begin{equation}
	\mathsf{A}=
	\begin{pmatrix}
		-1/2& 1/2& 1/2\\
		 1/2&-1/2& 1/2\\
		 1/2& 1/2&-1/2
	\end{pmatrix}.
	\label{eq:S_primitive_transform}
\end{equation}
For a point operation represented by $R_g$ in the conventional frame, its action in primitive coordinates is $W_g=\mathsf{A}^{-1}R_g\mathsf{A}$, and the corresponding momentum transformation is $\mathbf{k}\mapsto W_g^{-T}\mathbf{k}$.  A minimal set of generators for the $D_{4h}$ point group is
\begin{equation}
	R_{C_{4z}}=
	\begin{pmatrix}0&-1&0\\1&0&0\\0&0&1\end{pmatrix},
	\qquad
	R_{C_{2x}}=
	\begin{pmatrix}1&0&0\\0&-1&0\\0&0&-1\end{pmatrix},
	\qquad
	R_{\mathcal I}=\operatorname{diag}(-1,-1,-1).
	\label{eq:S_spatial_generators}
\end{equation}
The remaining operations used in the symmetry projection are generated from these matrices; for example,
$C_{2z}=C_{4z}^{2}$, $C_{2y}=C_{4z}C_{2x}C_{4z}^{-1}$,
$\sigma_h=\mathcal I C_{2z}$, and $S_{4z}=\mathcal I C_{4z}$.

In the local orbital subspaces, the nontrivial generator matrices are
\begin{equation}
	\rho^{\rm Mo}_{C_{4z}}=
	\begin{pmatrix}0&-1\\1&0\end{pmatrix},
	\qquad
	\rho^{\rm Mo}_{C_{2x}}=
	\begin{pmatrix}-1&0\\0&1\end{pmatrix},
	\qquad
	\rho^{\rm Th}_{C_{4z}}=-1,
	\qquad
	\rho^{\rm Th}_{C_{2x}}=1,
	\label{eq:S_orbital_generators}
\end{equation}
where the Mo matrices act on $(d_{xz},d_{yz})$ and the Th matrices act on $d_{x^2-y^2}$.  All retained $d$ orbitals are even under inversion.  The full orbital representation $U_g^{\rm orb}(\mathbf{k})$ additionally contains the permutation of the two Mo and two Th sites and the Bloch phase associated with any primitive-lattice translation.  Including spin,
\begin{equation}
	D_g(\mathbf{k})=U_g^{\rm spin}\otimes U_g^{\rm orb}(\mathbf{k}),
	\qquad
	U^{\rm spin}_{C_{4z}}=e^{-i\pi s_z/4},\quad
	U^{\rm spin}_{C_{2x}}=-i s_x,\quad
	U^{\rm spin}_{\mathcal I}=s_0.
	\label{eq:S_spinful_generators}
\end{equation}
Thus, it is sufficient to impose the generator constraints; the relations above guarantee all 16 operations of $D_{4h}$.  Together with Hermiticity and time-reversal symmetry, the constraints are
\begin{equation}
	D_g^{\dagger}(\mathbf{k})H(g\mathbf{k})D_g(\mathbf{k})=H(\mathbf{k}),
	\qquad
	\mathcal{T}H(\mathbf{k})\mathcal{T}^{-1}=H(-\mathbf{k}),
	\qquad
	\mathcal{T}=i s_y\mathcal{K}.
	\label{eq:S_SOC_constraints}
\end{equation}

We applied these constraints to the most general Mo--Th spin-flip ansatz constructed from $s_x$ and $s_y$.  For the lowest-order odd form factor $\sin(k_1+k_2)$, the symmetry projection leaves two independent invariant matrices. The resulting term is
\begin{equation}
	H_{\lambda_3}(\mathbf{k})
	=
	\lambda_3 \sin(k_1+k_2)\, \mathcal{M}_{3},
	\label{eq:S_Hlambda3}
\end{equation}
where, in the spin ordering of Eq.~\eqref{eq:S_primitive_basis_spinful},
\begin{equation}
	\mathcal M_3=
	\begin{pmatrix}0&A_3\\A_3^{\dagger}&0\end{pmatrix},
	\qquad
	A_3=
	\begin{pmatrix}
		0& 0& 0&0& 0& 0\\
		0& 0& 0&0& 1&-1\\
		0& 0& 0&0&-i& i\\
		0& 0& 0&0& 0& 0\\
		0& 1&-i&0& 0& 0\\
		0&-1& i&0& 0& 0
	\end{pmatrix}.
	\label{eq:S_M3}
\end{equation}
This explicitly shows that the $\lambda_3$ term couples the Mo and Th sectors and flips spin.

Physically, $H_{\lambda_1}$ describes the local atomic SOC on the Mo $d_{xz}/d_{yz}$ doublet, while $H_{\lambda_3}$ represents the symmetry-allowed Mo--Th interorbital SOC channel. The latter is essential for opening the SOC-induced gap in the inverted Mo--Th low-energy bands while preserving the crystalline symmetries and time-reversal symmetry.

\subsection{Parameterization}

\begin{figure}[H]
	\centering
	\includegraphics[width=0.8\linewidth]{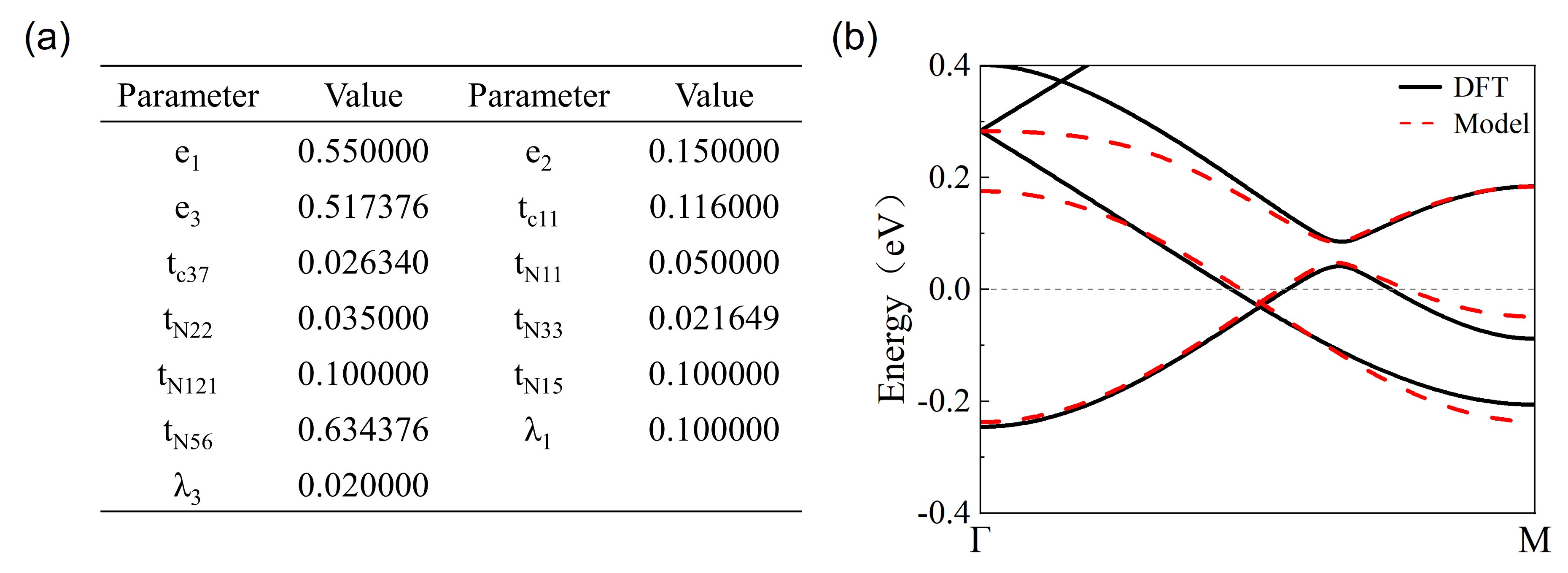}
	\caption{Parameters and fitting quality of the low-energy effective model. (a) Parameters of the effective tight-binding Hamiltonian in unit of eV. (b) Comparison between the DFT and effective-model band structures along the primitive-cell $\Gamma$--$M$ path. The black solid curves show the DFT bands including SOC, whereas the red dashed curves show the fitted low-energy effective model.}
	\label{fig:S_band_fitting}
\end{figure}

The model parameters were obtained by fitting the low-energy bands near the SOC-induced target gap. The resulting parameter set and fitting quality are summarized in Fig.~\ref{fig:S_band_fitting}(a) and (b), respectively. Here $e_i$ denote onsite energies, $t_N$ parameters denote in-plane hopping amplitudes, $t_c$ parameters denote inter-unit hopping amplitudes, and $\lambda_i$ denote SOC amplitudes. Along the primitive-cell $\Gamma$--$M$ path, the fitted model reproduces the DFT dispersion in the energy window relevant to the band inversion and the subsequent superconducting calculations. 

\subsection{Topology of the effective model}

\begin{figure}[htbp]
	\centering
	\includegraphics[width=\linewidth]{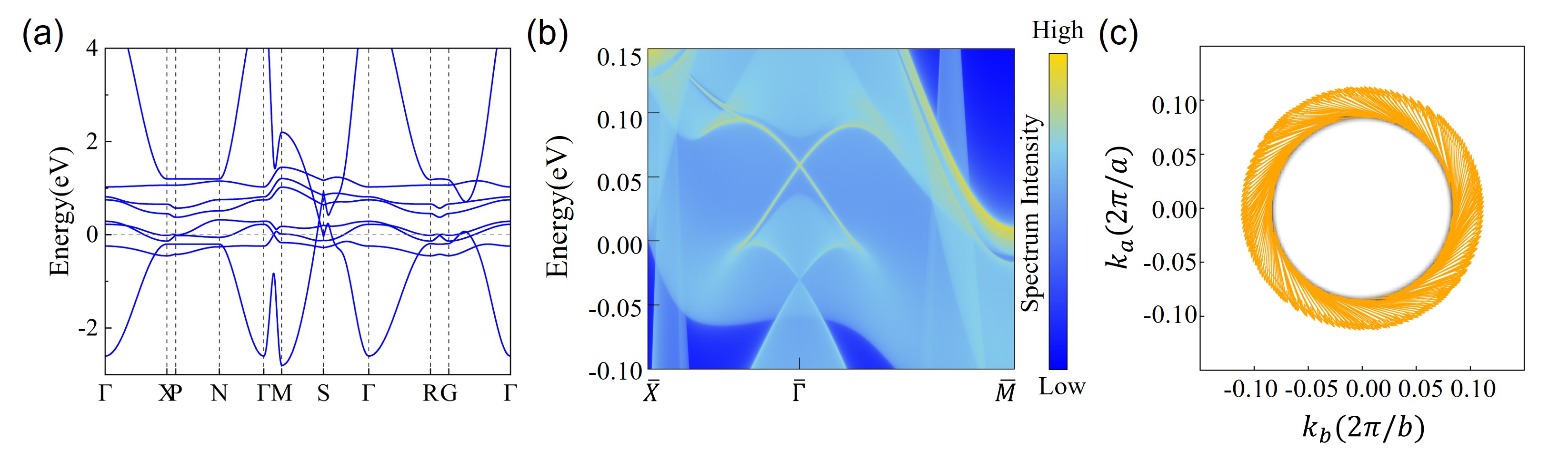}
	\caption{Topology of the orbital-extended effective model. (a) Bulk band structure after including the C-$s$ and C-$p_z$ orbitals. (b) (001) surface spectra with SOC. (c) In-plane spin texture at $0.04$~eV. The enlarged model restores the weak index $(0;111)$ of the DFT band manifold while preserving the low-energy Mo--Th band inversion and the spin-momentum-locked (001) surface state.}
	\label{fig:S_extended_topology}
\end{figure}

For the minimal Mo--Th orbital manifold introduced above, the Fu--Kane parity products give the strong topological index
\begin{equation}
	(\nu_0;\nu_1\nu_2\nu_3)=(1;000).
\end{equation}

This differs from the weak index $(0;111)$ obtained for the selected first-principles band manifold in Sec.~S2. The two indices classify \emph{different Hamiltonians and band manifolds}, and therefore should not be identified. In particular, a change of the strong $\mathbb Z_2$ index cannot result from merely appending decoupled topologically trivial orbitals. The extended model must be evaluated using its actual band ordering, hybridization, and chosen direct gap. Including C-$s$ and C-$p_z$ degrees of freedom together with their hybridizations gives

\begin{equation}
	(\nu_0;\nu_1\nu_2\nu_3)=(0;111),
\end{equation}
in agreement with the DFT result. As shown in Fig.~\ref{fig:S_extended_topology}, these additional orbitals do not modify the Mo--Th band inversion or the (001) surface states in the energy window relevant to the superconducting analysis. We therefore use the minimal model below, where it provides an economical representation of the target surface state and vortex bands, while the enlarged model establishes consistency with the bulk topological indices.

\section{S5. Phonons and Electron--Phonon Coupling}

\subsection{Computational details}

Phonon and electron--phonon coupling calculations were carried out using the \textsc{Quantum ESPRESSO} package~\cite{SM:Giannozzi2009} within the framework of density-functional perturbation theory (DFPT)~\cite{SM:Baroni2001}. To maintain the same exchange--correlation description as in the VASP calculations, the PBE functional was also employed in the \textsc{Quantum ESPRESSO} calculations. Ultrasoft pseudopotentials (USPPs) from the PSlibrary~\cite{SM:DalCorso2014} were used for all atomic species. For example, the Th pseudopotential was \texttt{Th.pbe-spfn-rrkjus\_psl.1.0.0.UPF}; in the PSlibrary naming convention, the \texttt{rrkjus} label denotes a Rappe--Rabe--Kaxiras--Joannopoulos ultrasoft pseudopotential. The kinetic-energy cutoffs for the wave functions and charge density were set to 80 and 960~Ry, respectively.

The electronic ground state used for the phonon calculations was obtained on an $8\times8\times8$ $k$-point mesh. Lattice-dynamical and electron--phonon matrix elements were evaluated on a $4\times4\times4$ phonon $q$-point mesh, while a denser $16\times16\times16$ electronic $k$-point mesh was employed in the evaluation of the Eliashberg spectral function $\alpha^2F(\omega)$ and the electron--phonon coupling strength. Because DFPT calculations for this multicomponent intergrowth structure are computationally demanding, the $4\times4\times4$ $q$ mesh was chosen as a practical compromise between Brillouin-zone sampling and computational cost. Accordingly, the electron--phonon results are used here primarily to establish a finite phonon-mediated pairing interaction and to provide a semiquantitative estimate of the superconducting transition temperature rather than a high-precision determination of $T_c$. As an internal cross-check, the electronic band structure and phonon dispersion obtained from the \textsc{Quantum ESPRESSO}-based calculations were compared with independent VASP-based results and found to be consistent.

\subsection{Electron--phonon coupling and transition temperature}

\begin{figure}[htbp]
	\centering
	\includegraphics[width=\linewidth]{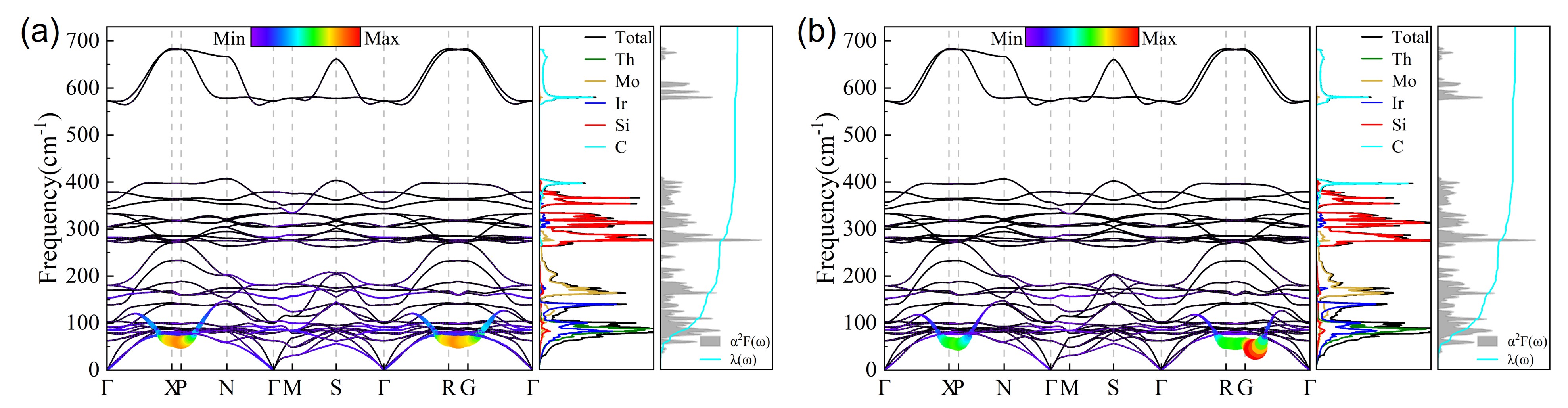}
	\caption{Calculated phonon properties and electron--phonon coupling of Th$_2$Mo$_2$Ir$_2$Si$_4$C for (a) the undoped system and (b) electron doping of $0.05$ electrons per primitive cell, for which the Fermi level is shifted into the topological gap. Each panel shows the phonon dispersion with mode-resolved phonon linewidths $\gamma_{\mathbf{q}\nu}$, the atom-resolved phonon density of states, the Eliashberg spectral function $\alpha^2F(\omega)$, and the accumulated electron--phonon coupling constant $\lambda(\omega)$. }
	\label{fig:S_epc}
\end{figure}

The isotropic Eliashberg coupling is characterized by
\begin{equation}
\lambda=2\int_0^\infty
\frac{\alpha^2F(\omega)}{\omega}\,d\omega,
\end{equation}
and
\begin{equation}
\omega_{\log}=
\exp\left[
\frac{2}{\lambda}
\int_0^\infty
\frac{\alpha^2F(\omega)}{\omega}
\ln\omega\,d\omega
\right].
\end{equation}
Using the Allen--Dynes-modified McMillan expression,
\begin{equation}
T_c=
\frac{\omega_{\log}}{1.2}
\exp\left[
-\frac{1.04(1+\lambda)}
{\lambda-\mu^*(1+0.62\lambda)}
\right],
\end{equation}
with $\mu^*=0.1$, the undoped calculation gives $\lambda=0.68752$ and $T_c=6.29$~K. For electron doping of $0.05$ electrons per primitive cell, we obtain $\lambda=0.69135$ and $T_c=6.12$~K. 
The comparable coupling strengths and finite estimated transition temperatures support the presence of a phonon-mediated pairing channel upon shifting the chemical potential into the target gap. These calculations do not, by themselves, determine the symmetry of the experimentally realized superconducting order parameter. The onsite spin-singlet pairing used in the BdG calculations is a modeling simplification.

\section{S6. Uniform and Vortex Superconductivity}

\subsection{Uniform superconducting state}

\begin{figure}[htbp]
	\centering
	\includegraphics[width=0.5\linewidth]{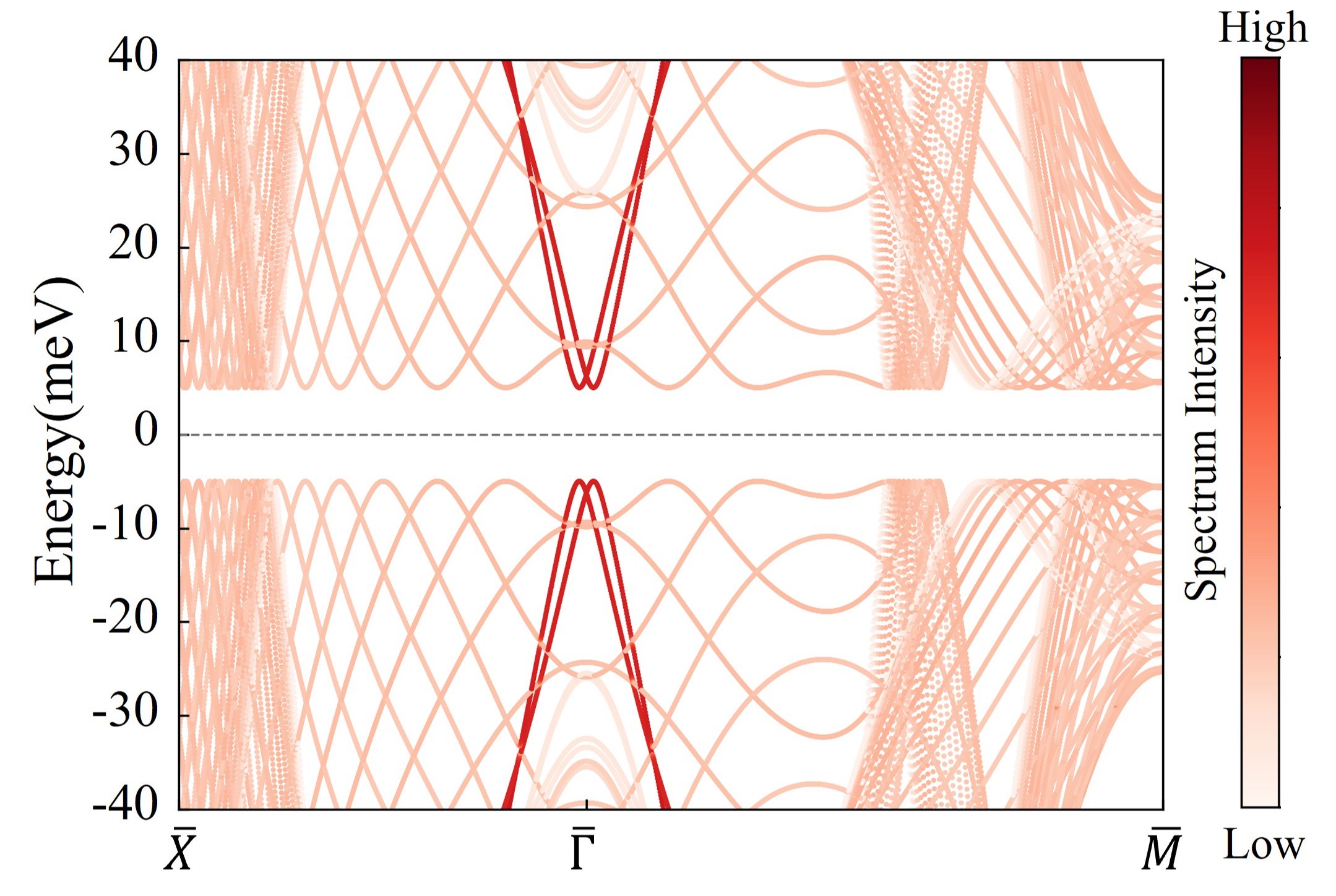}
	\caption{BdG spectrum of a (001) slab with uniform spin-singlet pairing at $\mu=65$~meV and $\Delta_0=5$~meV. The normal-state topological surface states are fully gapped by the superconducting pairing.}
	\label{fig:S_uniform_bdg}
\end{figure}

We first consider a uniform onsite spin-singlet pairing potential in a slab that is open along $[001]$ and periodic in the in-plane directions. In the Nambu basis, the slab Bogoliubov--de Gennes (BdG) Hamiltonian is
\begin{equation}
	\mathcal{H}_{\mathrm{BdG}}^{\mathrm{slab}}(\mathbf{k}_{\parallel})=
	\begin{pmatrix}
	H_{\mathrm{slab}}(\mathbf{k}_{\parallel})-\mu & \hat{\Delta}_0\\
	\hat{\Delta}_0^{\dagger} & -H_{\mathrm{slab}}^{*}(-\mathbf{k}_{\parallel})+\mu
	\end{pmatrix},
	\qquad
	\hat{\Delta}_0=\Delta_0 i s_y .
	\label{eq:S_uniform_bdg}
\end{equation}
The pairing term gaps both the bulk bands and the (001) topological surface state, as illustrated in Fig.~\ref{fig:S_uniform_bdg}.

\subsection{Vortex-line Hamiltonian and Majorana number}

For a straight vortex line parallel to the crystallographic $c$ axis, translational symmetry is retained only along the vortex direction. The corresponding BdG Hamiltonian at momentum $k_C$ is
\begin{equation}
	\mathcal{H}_{\mathrm{BdG}}(k_C)=
	\begin{pmatrix}
	H_0(k_C)-\mu & \hat{\Delta}(\mathbf{r})\\
	\hat{\Delta}^{\dagger}(\mathbf{r}) & -H_0^{*}(-k_C)+\mu
	\end{pmatrix},
	\label{eq:S_vortex_bdg}
\end{equation}
with
\begin{equation}
	\hat{\Delta}(\mathbf{r})=
	\Delta_0\tanh\!\left(\frac{|\mathbf{r}-\mathbf{r}_0|}{\xi}\right)
	e^{i\arg(\mathbf{r}-\mathbf{r}_0)} i s_y ,
	\label{eq:S_vortex_profile}
\end{equation}

where $\mathbf{r}_0$ is the vortex center and $\xi$ characterizes the vortex-core profile. For fixed vorticity, the \emph{one-dimensional family} $\mathcal H_{\mathrm{BdG}}(k_C)$ belongs to class D and satisfies $\mathcal C\mathcal H_{\mathrm{BdG}}(k_C)\mathcal C^{-1}=-\mathcal H_{\mathrm{BdG}}(-k_C)$, with particle--hole operator $\mathcal C=U_{\mathcal C}\mathcal K$. At the particle--hole-invariant momenta $k_C=0,\pi$, the matrices

\begin{equation}
	B(k_C)=\mathcal{H}_{\mathrm{BdG}}(k_C)U_{\mathcal{C}},
	\qquad k_C=0,\pi,
\end{equation}
are antisymmetric. The endpoint Pfaffian phases and the Majorana number are
\begin{equation}
	p_{k_C}=\frac{\operatorname{Pf}B(k_C)}{|\operatorname{Pf}B(k_C)|},
	\qquad
	\mathcal Q_{v}=\operatorname{sgn}\operatorname{Re}\!\left(\frac{p_0}{p_{\pi}}\right)\in\{+1,-1\}.
	\label{eq:S_majorana_number}
\end{equation}

Here the endpoint phases are compared in one fixed particle--hole-compatible convention. In a corresponding real Majorana basis, this expression reduces to the familiar sign of the product of the two endpoint Pfaffians, as used in the main text,
\begin{equation}
\mathcal{Q}_v =
\operatorname{sgn}\!\left[
\operatorname{Pf} B(0)\,
\operatorname{Pf} B(\pi)
\right],
\label{eq:pfaffian-11}
\end{equation}
When $E_{\mathrm{vg}}=\min_{k_C,n}|E_n(k_C)|>0$, $\mathcal Q_v=-1$ identifies the nontrivial one-dimensional class-D phase with Majorana end modes, whereas $\mathcal Q_v=+1$ is trivial~\cite{SM:Kitaev2001}. The endpoint Pfaffian signs can still be evaluated in a gapless nodal or hybrid regime, but they do not label a \emph{globally gapped} one-dimensional topological phase.

\subsection{$C_4$ eigensectors and symmetry-protected vortex nodes}

\begin{figure}[htbp]
	\centering
	\includegraphics[width=0.7\linewidth]{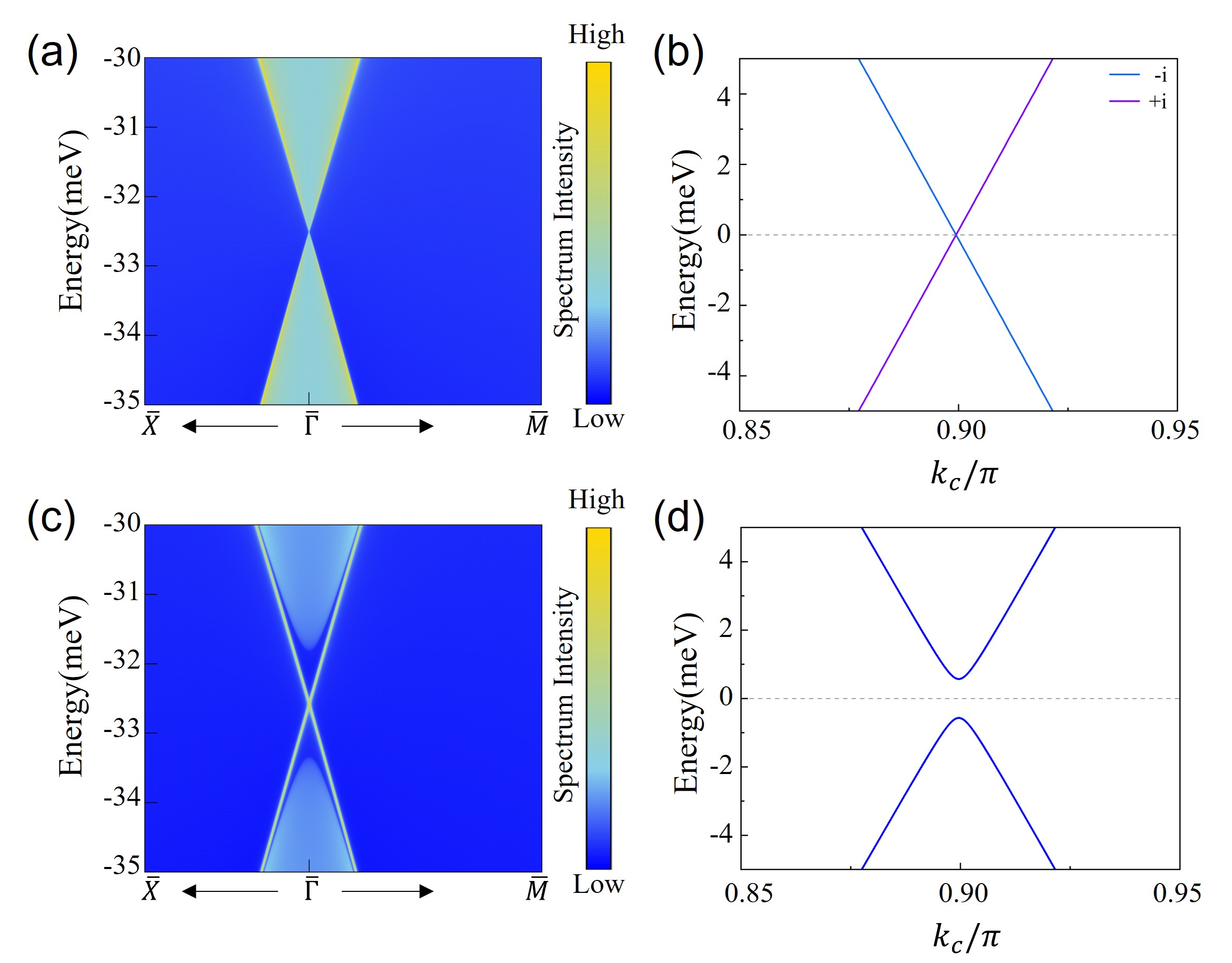}
	\caption{Effect of $C_4$ breaking on the normal-state surface spectrum and vortex-line BdG spectrum. (a) (001)-surface-projected spectral function in the $C_4$-symmetric limit, showing the surface projection of the bulk Dirac cone at $\bar{\Gamma}$. (b) $C_4$-resolved vortex-line spectrum at $\mu=-75$~meV. The $+i$ (purple) and $-i$ (blue) branches belong to distinct rotation eigensectors and therefore cross without hybridizing, producing a symmetry-protected node. (c) Corresponding surface spectral function after introducing the orbital-nematic perturbation $\delta_{C_4}=5$~meV, i.e., an orbital splitting $\delta=2\delta_{C_4}=10$~meV. (d) Corresponding vortex-line spectrum after $C_4$ breaking.}
	\label{fig:S_c4_spectra}
\end{figure}

When the vortex is centered on a crystallographic fourfold axis, the BdG Hamiltonian commutes with the gauge-dressed rotation operator
\begin{equation}
	\widetilde{C}_4=e^{-i\pi\tau_z/4}
	\begin{pmatrix}
	U_{C_4} & 0\\
	0 & U_{C_4}^{*}
	\end{pmatrix},
	\qquad
	[\mathcal{H}_{\mathrm{BdG}}(k_C),\widetilde{C}_4]=0,
	\label{eq:S_bdg_c4}
\end{equation}
where $\tau_z$ acts in Nambu space and the prefactor compensates the phase winding of the vortex order parameter. The BdG eigenstates can therefore be classified by the four eigenvalues $C_4=+1,-1,+i,-i$. For an isolated state $|u_n(k_C)\rangle$, the label is obtained from $\langle u_n|\widetilde{C}_4|u_n\rangle$; within a degenerate subspace, it is obtained by diagonalizing the projected operator $P\widetilde{C}_4P$. Bands belonging to distinct $C_4$ eigensectors cannot hybridize. In particular, crossings between the $+i$ and $-i$ sectors remain gapless and generate the nodal portions of the vortex phase diagram, as illustrated in Fig.~\ref{fig:S_c4_spectra}(b).

This protection can be removed by the orbital-nematic perturbation
\begin{equation}
	V_{C_4}=\delta_{C_4}
	\sum_{a\in\mathcal{I}_{\mathrm{Mo}}}
	\sum_{\sigma=\uparrow,\downarrow}
	\left(n_{a,xz,\sigma}-n_{a,yz,\sigma}\right),
	\qquad
	\delta\mathcal{H}_{\mathrm{BdG}}=
	\begin{pmatrix}V_{C_4}&0\\0&-V_{C_4}^{*}\end{pmatrix},
	\label{eq:S_c4_breaking}
\end{equation}
Here $\mathcal{I}_{\mathrm{Mo}}$ denotes the set of Mo sites, $\sigma$ labels the spin projection, and $\delta_{C_4}$ shifts the Mo $d_{xz}$ and $d_{yz}$ orbitals in opposite directions; hence the corresponding orbital splitting is $\delta=2\delta_{C_4}$. Figure~\ref{fig:S_c4_spectra} compares the normal-state surface spectrum and the vortex spectrum before and after this perturbation. In the $C_4$-symmetric limit, the projection of the bulk Dirac cone onto the (001) surface remains gapless [Fig.~\ref{fig:S_c4_spectra}(a)], while the $+i$ and $-i$ vortex branches cross at zero energy for $\mu=-75$~meV [Fig.~\ref{fig:S_c4_spectra}(b)]. For $\delta_{C_4}=5$~meV, corresponding to an orbital splitting $\delta=10$~meV, the bulk Dirac crossing is gapped and a nontrivial Dirac-like surface state appears inside the resulting gap [Fig.~\ref{fig:S_c4_spectra}(c)]. 
The same perturbation mixes the two vortex rotation sectors and converts their protected crossing into an avoided crossing [Fig.~\ref{fig:S_c4_spectra}(d)]. This removes the illustrated symmetry-protected node and is consistent with the gapped surface-Kitaev region identified from the full momentum-dependent vortex calculation. Displacing the vortex center away from the fourfold axis likewise breaks the global vortex $C_4$ symmetry and opens a finite, although possibly very small, gap at the former node.

\subsection{Orbital origin of the orbital Kitaev phase}

The low-chemical-potential topological interval does not originate from the (001) topological surface state. For a Th-centered vortex, the order parameter vanishes at the Th site located at the vortex core. Neglecting intersite hybridization, the local core Hamiltonian is
\begin{equation}
	\mathcal{H}_{\mathrm{core}}^{\mathrm{Th}}
	\simeq (\varepsilon_{\mathrm{Th}}-\mu)\tau_z
	+\operatorname{Re}\Delta(\mathbf{r}_0)\tau_x
	-\operatorname{Im}\Delta(\mathbf{r}_0)\tau_y,
\end{equation}
and $\Delta(\mathbf{r}_0)=0$ gives
\begin{equation}
	E_{\mathrm{core}}=\pm|\varepsilon_{\mathrm{Th}}-\mu|.
	\label{eq:S_th_core_level}
\end{equation}
Interlayer hopping and hybridization disperse this local Th-derived level into a one-dimensional vortex band. Its successive band inversions at $k_C=0$ and $\pi$ delimit a gapped interval with $\mathcal Q_v=-1$. We refer to this phase as the \emph{orbital Kitaev phase}: its topology is generated by the Th-derived vortex-core band rather than by a proximitized topological surface state. By contrast, the high-chemical-potential \emph{surface Kitaev phase} descends from the superconducting gap opened in the (001) topological surface states.

\begin{figure}[t]
	\centering
	\includegraphics[width=0.75\linewidth]{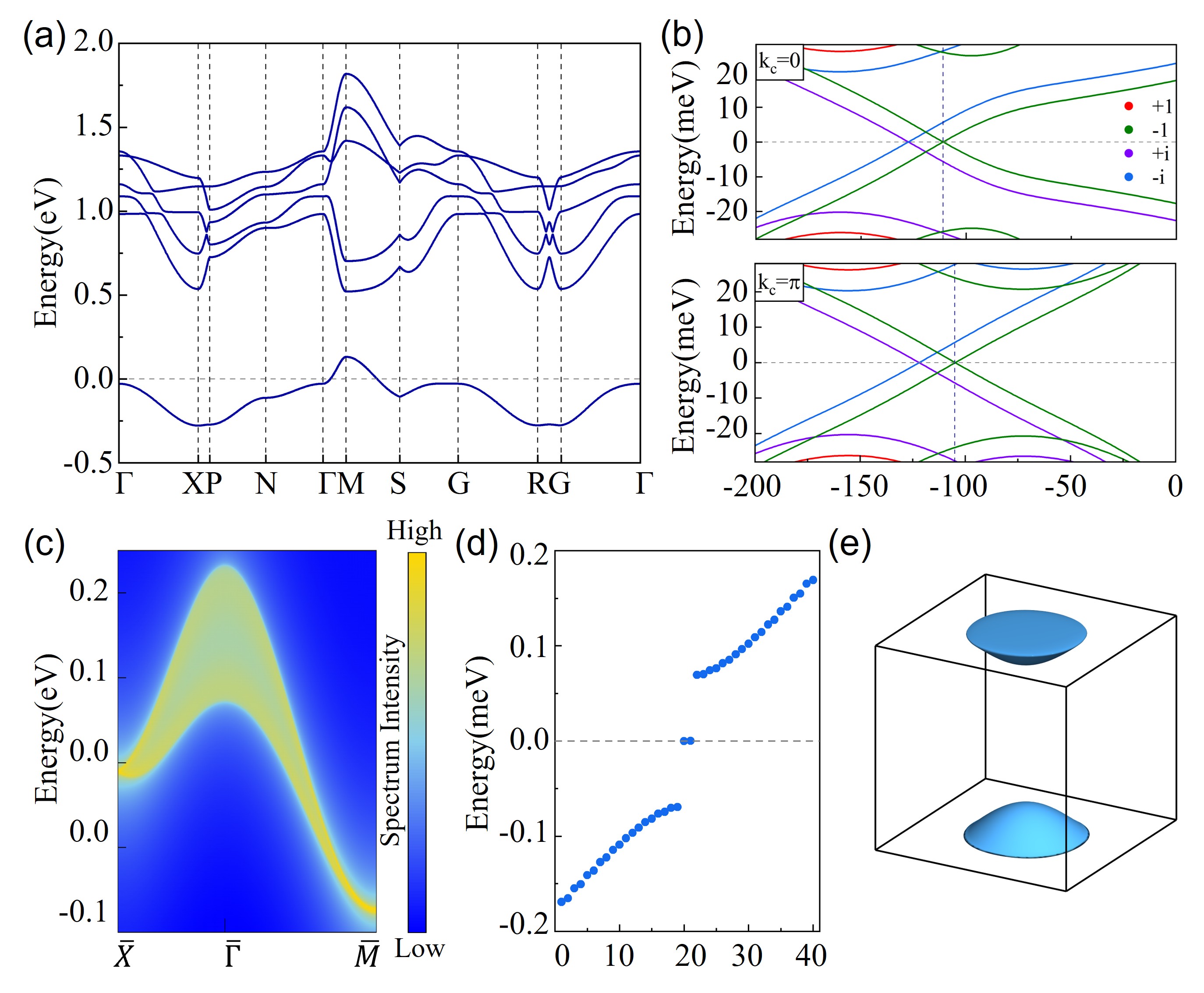}
	\caption{Orbital Kitaev phase without normal-state topological surface states. (a) Normal-state band structure of the modified effective model with $t_{c37}=0.01$~eV, $e_1=e_2=1$~eV, and $\lambda_3=0.08$~eV. The Mo-derived bands are shifted away from the Fermi level, leaving an isolated Th-derived low-energy band. (b) $C_4$-resolved vortex-line BdG spectra at $k_C=0$ (upper) and $k_C=\pi$ (lower) in the presence of the $c$-axis Zeeman term $H_Z=B_z\sigma_z$. Here $B_z=2.5$~meV, corresponding to a spin splitting $2B_z=5$~meV. Red, green, purple, and blue denote the $C_4=+1,-1,+i$, and $-i$ sectors, respectively, and the vertical dashed lines delimit the fully gapped orbital-Kitaev interval. (c) (001)-surface spectral function for the same modified model. (d) Low-energy eigenvalues of the three-dimensionally finite system with open boundaries in all three directions at $\mu=-108$~meV. (e) Real-space probability density of the corresponding Majorana subspace, localized at the two terminations of the vortex line.}
	\label{fig:S_single_th_orbital_kitaev}
\end{figure}

Figure~\ref{fig:S_single_th_orbital_kitaev} provides a direct control calculation that isolates the orbital mechanism from the normal-state topological surface-state mechanism. We set $t_{c37}=0.01$~eV, raise the Mo-orbital onsite energies to $e_1=e_2=1$~eV, and enhance the Mo--Th spin-orbit coupling to $\lambda_3=0.08$~eV, while keeping all other effective-model parameters unchanged. As shown in Fig.~\ref{fig:S_single_th_orbital_kitaev}(a), the Mo-derived bands are thereby displaced far from the Fermi level and only an isolated Th-derived band remains in the relevant low-energy window. The Mo--Th band inversion responsible for the (001) topological surface states of the full model is therefore absent (as shown in Fig.~\ref{fig:S_single_th_orbital_kitaev}(c)). This modified Hamiltonian consequently provides a stringent test of whether the low-chemical-potential Kitaev phase can arise from the Th-derived vortex-core orbital.

To separate the $C_4=-1$ Kitaev interval from the zero-energy crossings in the $C_4=+i$ and $-i$ sectors, we further include a $c$-axis Zeeman term $H_Z=B_z\sigma_z$. In this convention the two eigenvalues of $\sigma_z$ are $\pm1$. Hence $B_z=2.5$~meV shifts the two spin sectors by $\pm2.5$~meV and produces a total spin splitting of $2B_z=5$~meV. The perturbation breaks time-reversal symmetry but preserves rotation about the $c$ axis. Note that the Zeeman term is introduced as a theoretical control perturbation to separate coincident vortex-band inversions, rather than as a quantitative prediction for an experimentally accessible magnetic field in this compound.

The $k_C=0$ and $k_C=\pi$ spectra in Fig.~\ref{fig:S_single_th_orbital_kitaev}(b) show that the two zero-energy inversions associated with the $C_4=-1$ vortex branch delimit a fully gapped interval for $-111~\mathrm{meV}<\mu<-104~\mathrm{meV}$. The $C_4=\pm i$ crossings are displaced outside this interval, so no nodal vortex channel remains there. We have numerically checked that the low-energy spectrum is fully gapped in this chemical potential interval. The resulting one-dimensional class-D vortex Hamiltonian has Majorana number $\mathcal Q_v=-1$ and realizes a clean orbital Kitaev phase. At the representative chemical potential $\mu=-108$~meV, the spectrum of the system with open boundaries in all three directions contains an isolated particle-hole-related pair of near-zero-energy states [Fig.~\ref{fig:S_single_th_orbital_kitaev}(d)]. Their combined probability density is concentrated at the two opposite ends of the vortex line [Fig.~\ref{fig:S_single_th_orbital_kitaev}(e)], as expected for Majorana end modes. Because the (001) surface calculation in Fig.~\ref{fig:S_single_th_orbital_kitaev}(c) explicitly excludes a normal-state topological surface state in this energy window, these end modes cannot originate from a proximitized surface Dirac cone. They instead arise from the Th-derived vortex-core orbital band, demonstrating that a pre-existing topological surface state is not required for the orbital Kitaev phase.


\subsection{Effective Kitaev description of the core-position transition}
The low-energy spectra shown in Fig.~3(d)--(f) of the main text can be fitted by a single-channel, nearest-neighbor Kitaev-type Hamiltonian,
\begin{equation}
 H_{\mathrm{core}}^{\mathrm{eff}}(k_C;\mathbf r_0)=
 \big[m(\mathbf r_0)+2t_z(\mathbf r_0)(1-\cos k_C)\big]\tau_z
 +\Delta_p(\mathbf r_0)\sin k_C\,\tau_y.
 \label{eq:S_effective_kitaev}
\end{equation}
Here $m$ is the effective mass at $k_C=0$, $t_z$ describes the longitudinal dispersion, and $\Delta_p$ parametrizes an odd-momentum pairing induced in the projected core channel. The latter does not imply odd-parity pairing in the original bulk superconductor. For this effective Hamiltonian, the endpoint masses and the gapped-phase invariant are
\begin{equation}
 M_0=m,\qquad M_{\pi}=m+4t_z,\qquad
 \mathcal M_{\mathrm{eff}}=\operatorname{sgn}\big[m(m+4t_z)\big].
 \label{eq:S_effective_invariant}
\end{equation}

The fitted mass changes sign while $M_{\pi}>0$ throughout the three configurations. Hence the effective model closes at $k_C=0$ and switches its class-D invariant. The simultaneous change in $\Delta_p$ indicates a reconstruction of the low-energy anomalous coupling as the vortex core moves. These parameters describe selected low-energy BdG branches. The topological diagnosis of the material-based vortex comes independently from its complete $k_C$ minigap and full-model Pfaffian calculation.

\begin{table}[htbp]
	\vspace{0.2em}
	\REVcolor
	\centering
	\caption{Fitted effective Kitaev-chain parameters (meV) for three representative vortex-core configurations. The endpoint mass $M_{\pi}$ is calculated from the fit parameters. The label ``critical'' refers to the gap closing at $k_C=0$ of the effective model.}
	\begin{tabular}{lrrrrc}
		\toprule
		Configuration & $m$ & $t_z$ & $\Delta_p$ & $M_{\pi}$ & $\mathcal M_{\mathrm{eff}}$ \\
		\midrule
		O-Kitaev & $-6.636$ & $6.383$ & $0.153$ & $18.896$ & $-1$ \\
		Critical & $0.000$ & $4.558$ & $1.457$ & $18.232$ & gapless \\
		Trivial & $2.490$ & $4.493$ & $5.499$ & $20.462$ & $+1$ \\
		\bottomrule
	\end{tabular}
	\label{tab:S_effective_kitaev_fit}
\end{table}

\subsection{Finite-vortex spectra in the nodal and hybrid regimes}

\begin{figure}[htbp]
	\centering
	\includegraphics[width=0.7\linewidth]{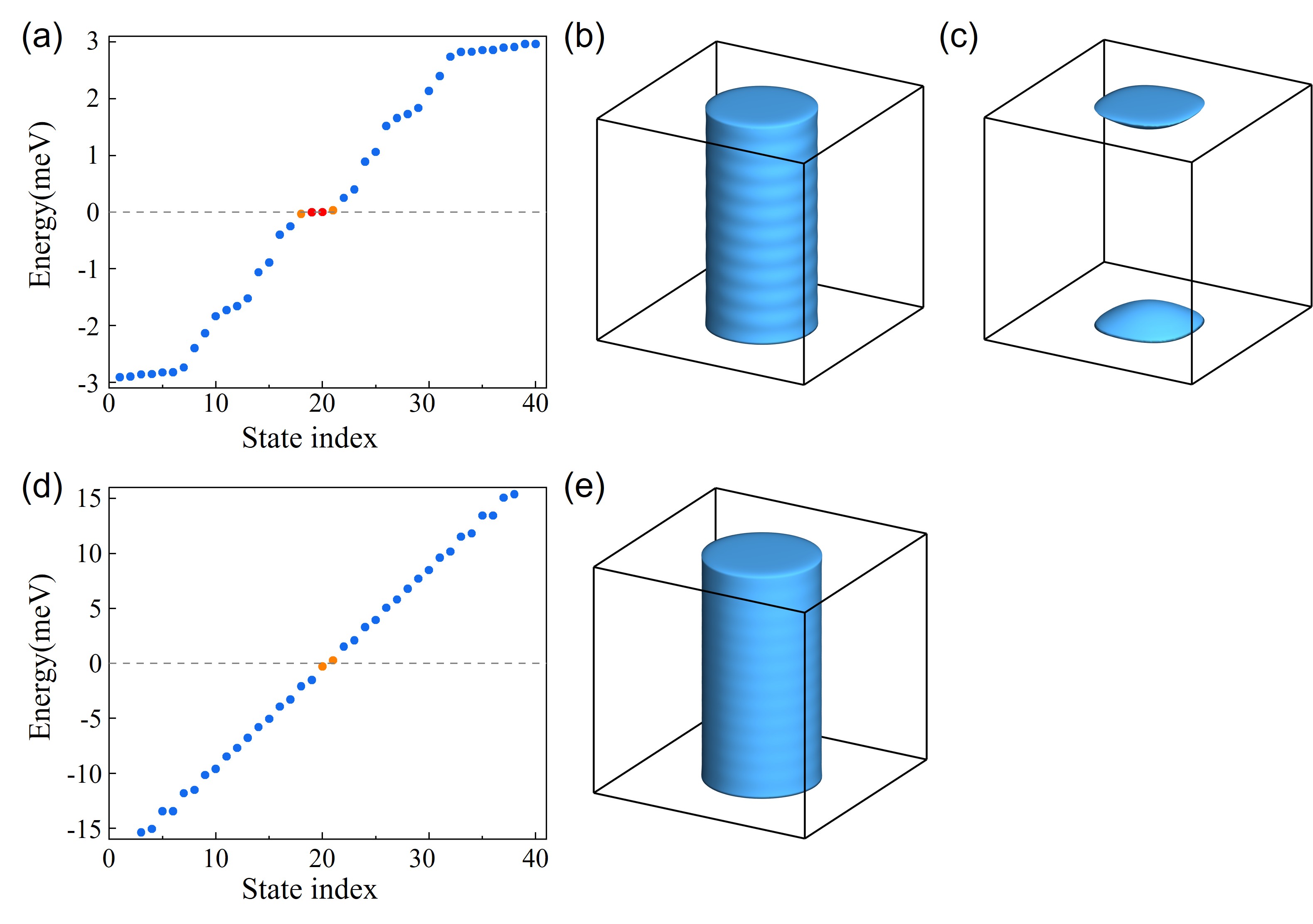}
	\caption{Finite-vortex BdG spectra and real-space probability densities in the hybrid and nodal regimes. (a) Low-energy BdG eigenvalues at $\mu=-25$~meV in the hybrid phase. The orange markers select the near-zero nodal-channel eigenstates whose probability density is shown in (b), whereas the red markers select the Majorana zero-mode subspace whose probability density is shown in (c). (d) Low-energy BdG eigenvalues at $\mu=-100$~meV in the nodal phase; the orange markers select the near-zero nodal-channel eigenstates whose probability density is shown in (e). }
	\label{fig:S_finite_vortex}
\end{figure}

To complement the orbital- and surface-Kitaev results presented in the main text, Fig.~\ref{fig:S_finite_vortex} resolves the low-energy finite-vortex eigenstates in the hybrid and nodal regimes. At $\mu=-25$~meV, the hybrid phase contains both a line-delocalized near-zero state associated with the nodal vortex channel [orange markers in Fig.~\ref{fig:S_finite_vortex}(a) and the corresponding probability density in Fig.~\ref{fig:S_finite_vortex}(b)] and a separate Majorana zero-mode subspace localized at the two vortex terminations [red markers in Fig.~\ref{fig:S_finite_vortex}(a) and the corresponding probability density in Fig.~\ref{fig:S_finite_vortex}(c)]. At $\mu=-100$~meV, the selected near-zero state [orange markers in Fig.~\ref{fig:S_finite_vortex}(d)] remains extended throughout the vortex line [Fig.~\ref{fig:S_finite_vortex}(e)], as expected for a nodal vortex channel. The coexistence of line-extended and end-localized low-energy states in the hybrid regime distinguishes it from a fully gapped Kitaev phase.

\subsection{Spin-resolved Majorana line densities}

We further characterize the Majorana end modes through their spin-resolved
in-plane profiles.  For a normalized Majorana BdG eigenstate
$\Phi=(u,v)^{T}$, we use the electron component $u$.  In the $s_z$ basis,
the electron spinor on orbital $m$ in the $c$-axis layer $\ell$ is written
as $u_{m\ell}(x,y)=(u_{\uparrow,m\ell},u_{\downarrow,m\ell})^{T}$.  After
summing over the transverse coordinate $y$ and all orbitals, the line
densities with spin parallel or antiparallel to the $x$, $y$, and $z$ axes
are obtained directly as
\begin{align}
	\rho_{\pm x}^{(\ell)}(x)
	&=\frac{1}{2}\sum_{y,m}
	\left|u_{\uparrow,m\ell}(x,y)
	\pm u_{\downarrow,m\ell}(x,y)\right|^{2}, \\
	\rho_{\pm y}^{(\ell)}(x)
	&=\frac{1}{2}\sum_{y,m}
	\left|u_{\uparrow,m\ell}(x,y)
	\mp i u_{\downarrow,m\ell}(x,y)\right|^{2}, \\
	\rho_{\uparrow z}^{(\ell)}(x)
	&=\sum_{y,m}\left|u_{\uparrow,m\ell}(x,y)\right|^{2},\qquad
	\rho_{\downarrow z}^{(\ell)}(x)
	=\sum_{y,m}\left|u_{\downarrow,m\ell}(x,y)\right|^{2}.
	\label{eq:S_spin_projected_line_density}
\end{align}
Here the upper and lower signs correspond to the positive and negative spin projections, respectively.
The Majorana wave functions are normalized globally. 

\begin{figure}[htbp]
	\centering
	\includegraphics[width=0.7\linewidth]{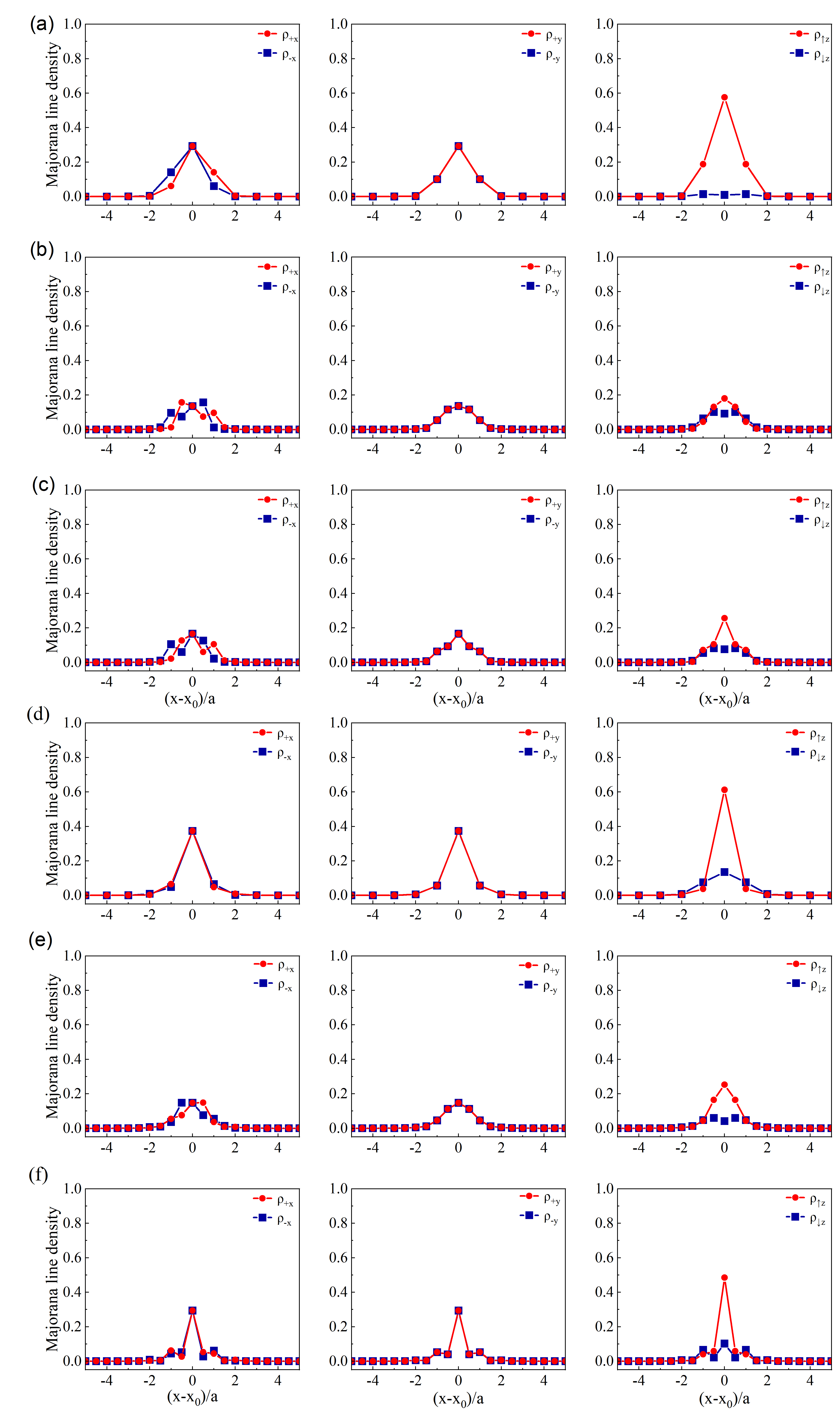}
	\caption{Spin-projected Majorana line densities in the surface and
	orbital Kitaev phases.  The three columns show the projections along the
	$x$, $y$, and $z$ spin axes, respectively.  Red circles and blue squares
	denote the positive and negative spin projections; in the third column
	they correspond to $\rho_{\uparrow z}$ and $\rho_{\downarrow z}$.
	(a--c) Surface Kitaev phase at $\mu=60$~meV: (a) outermost surface
	layer, (b) second-outermost surface layer, and (c) sum of the two layers.
	(d--f) Orbital Kitaev phase at $\mu=-260$~meV: (d) outermost surface
	layer, (e) second-outermost surface layer, and (f) sum of the two layers.
	The horizontal coordinate is measured relative to the vortex center
	$x_0$.}
	\label{fig:S_spin_density}
\end{figure}

Figure~\ref{fig:S_spin_density} shows that the Majorana end modes in both
Kitaev phases are sharply localized around the in-plane vortex position.
The pronounced imbalance between $\rho_{\uparrow z}$ and
$\rho_{\downarrow z}$ reveals a strong out-of-plane spin polarization,
whereas the $x$- and $y$-projected components are much more nearly
balanced. 
The finite weight in the second-outermost layer also resolves the penetration of the end mode beneath the surface. Because these quantities are electron-component projections of the BdG wave functions, they should be interpreted as near-surface spin-resolved spectral weights rather than tunneling conductances; an experimental comparison would additionally depend on orbital- and depth-dependent tunneling matrix elements.
\

\subsection{Robustness of the vortex BdG spectra}
\begin{figure}[H]
	\centering
	\includegraphics[width=0.9\linewidth]{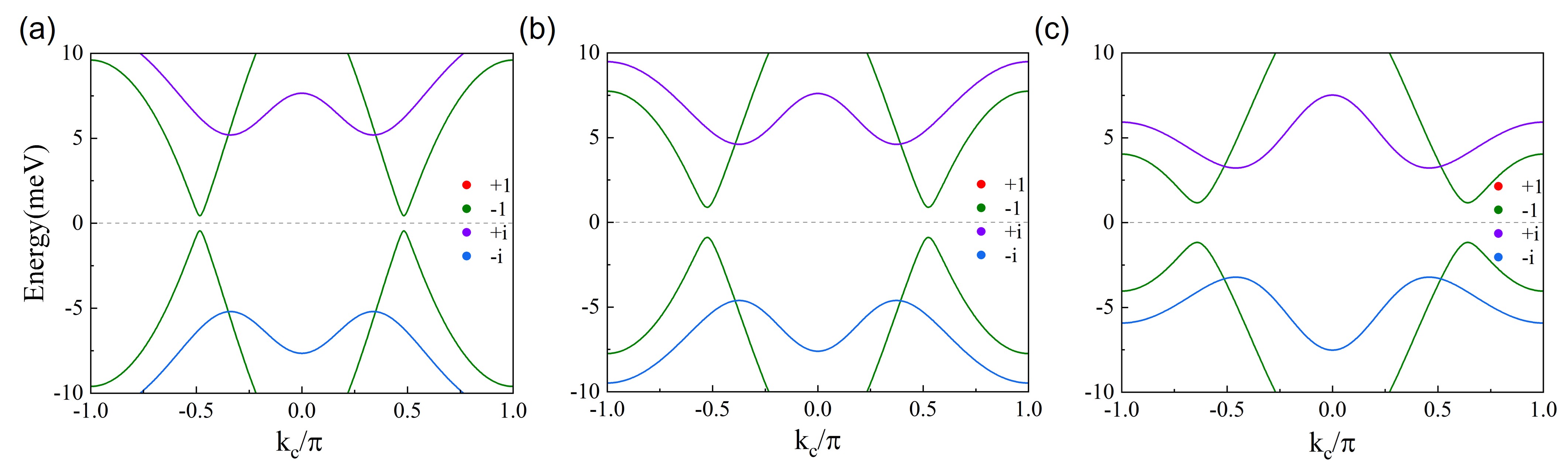}
	\caption{$C_4$-resolved vortex-line BdG spectra at $\mu=-280$~meV for
	enhanced Mo--Th coupling SOC.  (a--c) $\lambda_3=0.04$, $0.08$, and
	$0.12$~eV, corresponding to two, four, and six times the value in the
	original effective model, respectively.  Red, green, purple, and blue
	curves denote the $C_4=+1$, $-1$, $+i$, and $-i$ eigensectors.}
	\label{fig:S_lambda3_robustness}
\end{figure}

At $\mu=-280$~meV, the lowest excitation gap of the orbital Kitaev phase is controlled by the pair of branches in the
$C_4=-1$ eigensector. As shown in Fig.~\ref{fig:S_lambda3_robustness}, increasing the Mo--Th coupling SOC amplitude $\lambda_3$ progressively enlarges
this gap without inducing an intervening gap closing over the parameter range examined. We have checked that the global minigap and Pfaffian invariant across the SOC scan are robust. The orbital Kitaev phase
is therefore adiabatically preserved, while its spectral protection against weak perturbations and finite-size effects is
enhanced by stronger Mo--Th SOC.

\begin{figure}[hbp]
	\centering
	\includegraphics[width=\textwidth]{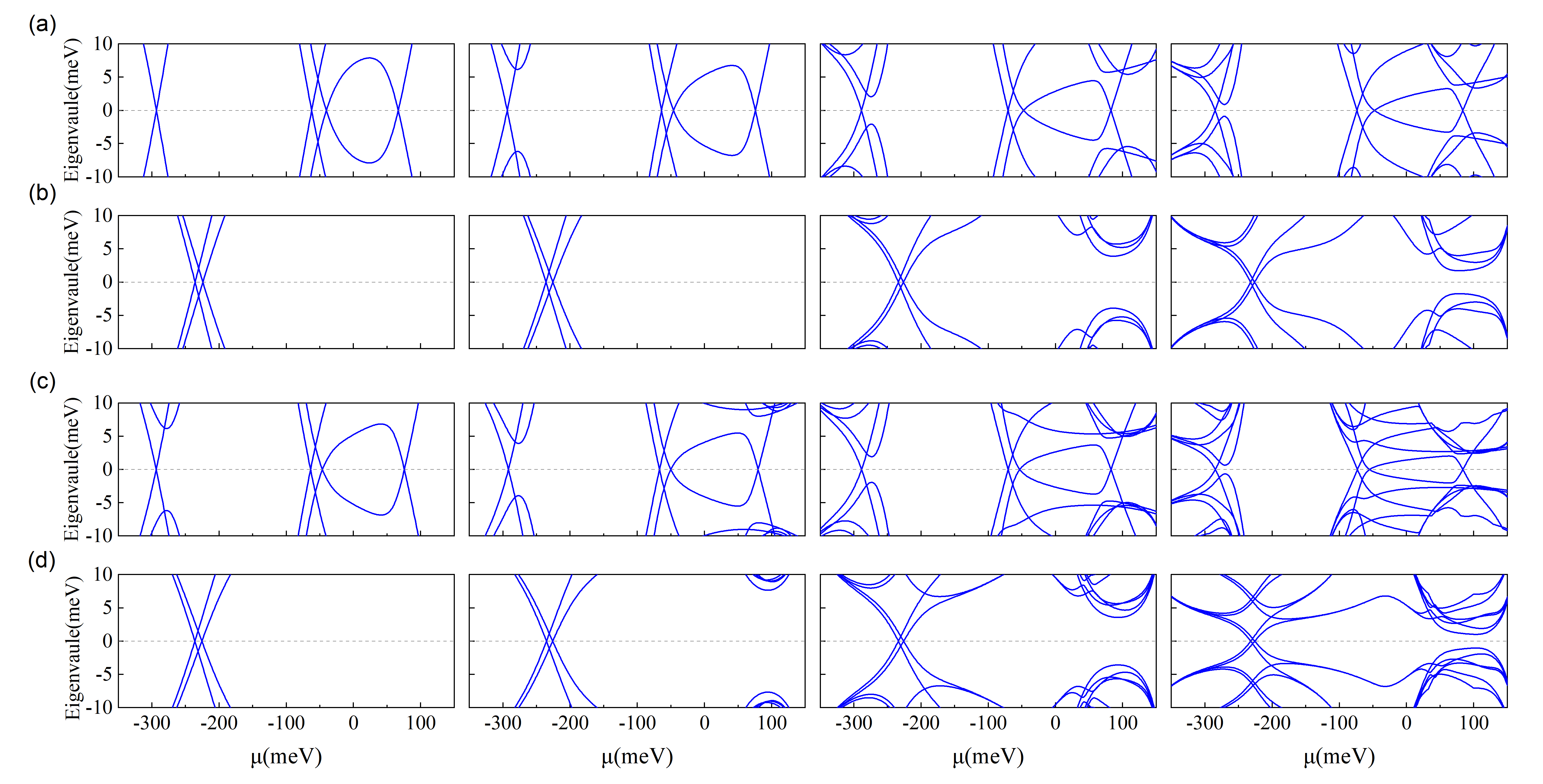}
	\caption{
		(a,b) BdG eigenvalues as functions of the chemical potential
		$\mu$ at $k_{c}=0$ and $k_{c}=\pi$, respectively, calculated
		with $\Delta_0=100~\mathrm{meV}$ and
		$\xi/a=1$, $1.5$, $3$, and $4.5$ from left to right.
		(c,d) Corresponding spectra at $k_{c}=0$ and $k_{c}=\pi$,
		respectively, calculated with $\xi=1.5a$ and
		$\Delta_0=100$, $75$, $50$, and $30~\mathrm{meV}$ from left
		to right. The horizontal dashed lines indicate zero energy.
	}
	\label{fig:S_parameter_robustness}
\end{figure}

The pairing amplitudes and vortex-core radii employed in these finite-size calculations are enlarged model parameters. To compare scales, one may invoke the clean-limit BCS estimate
\begin{equation}
 \xi_0\Delta_{\mathrm{BCS}}\simeq\frac{\hbar v_F}{\pi},
 \label{eq:S_clean_limit_relation}
\end{equation}
which is a rough reference scale rather than a prescription for mapping the microscopic BdG vortex profile from one parameter set to another. Using the reported gap estimate $\Delta_{\mathrm{exp}}=0.61$~meV and coherence length $\xi_{\mathrm{exp}}=18.6$~nm $\simeq44.05a$~\cite{SM:LiJACS2025}, keeping the product $\xi\Delta$ fixed would formally give $\xi/a\simeq0.90$, $0.54$, $0.36$, and $0.27$ at $\Delta_0=30$, $50$, $75$, and $100$~meV, respectively. These are rescaled estimates, not measured core sizes for enhanced pairing, and they do not establish a continuous, gapped path between the numerical and experimental Hamiltonians.

As shown in Fig. ~\ref{fig:S_parameter_robustness}, the characteristic low-energy band inversions persist upon increasing the vortex-core size at fixed pairing amplitude [Figs. ~\ref{fig:S_parameter_robustness}(a) and ~\ref{fig:S_parameter_robustness}(b)] and upon reducing the pairing amplitude at fixed core size [Figs. ~\ref{fig:S_parameter_robustness}(c) and ~\ref{fig:S_parameter_robustness}(d)]. In particular, the characteristic phase structure is retained down to $\Delta_0=30~\mathrm{meV}$ and $\xi=1.5a$, despite the increasingly dense low-energy spectrum. These results provide a limited test of the robustness of the calculated vortex phases against variations in the pairing amplitude and core size.
Note that, the displayed spectra are taken at the particle--hole-invariant momenta $k_C=0$ and $\pi$; we also checked that there are no nodes at intermediate $k_C$. 
Increasing $\Delta_0$ or decreasing $\xi$
separates the low-energy bands more clearly without introducing
additional gap closings. We therefore use Fig.~\ref{fig:S_parameter_robustness} as a limited parameter-sensitivity test, not as proof of an adiabatic connection to experimental scales or of zero-mode observability in the compound. The orbital-derived Kitaev mechanism and position-driven transition are established for the material-based multiorbital model and their quantitative realization needs future experimental verification.

\clearpage
\let\ThMoOriginalLabel\label
\renewcommand{\label}[1]{\ifstrequal{#1}{LastBibItem}{\ThMoOriginalLabel{SM-LastBibItem}}{\ThMoOriginalLabel{#1}}}

\end{document}